\documentclass[a4paper]{iopart}
\pdfoutput=1
\usepackage{hyperref}
\usepackage{textcomp}
\usepackage{amsmath}
\usepackage{amssymb}
\usepackage{graphicx} 
\usepackage{pgf}
\newcounter{defcounter}
\newenvironment{myequation}{%
\addtocounter{equation}{-1}
\refstepcounter{defcounter}
\renewcommand\theequation{C.\thedefcounter}
\begin{equation}}
{\end{equation}}

\newcommand{\ket}[1]{\left|#1\right\rangle}
\newcommand{\bra}[1]{\left\langle #1\right|}
\newcommand{\ud}{\mathrm{d}}
\newcommand{\tf}{t_{\mathrm{f}}}

\newcommand{\mean}[1]{\left\langle #1\right\rangle}

\newcommand{\nep}{\textrm{e}}

\newcommand{\Ham}{\widehat{H}}

\def\op#1{\widehat{#1}}
\def\ops#1{\hat{#1}}
\newcommand{\opc}[1]{{\hat{c}^{\protect\phantom \dagger}}_{#1}}
\newcommand{\opcdag}[1]{{\hat{c}^{\dagger}}_{#1}}

\newcommand{\opgammadag}[1]{{\hat{\gamma}^{\dagger}}_{#1}}

\begin{document}

\title{Asymptotic work statistics of periodically driven Ising chains}

\author{Angelo Russomanno$^{1,2,3}$, Shraddha Sharma$^4$, Amit Dutta$^4$, Giuseppe E. Santoro$^{1,2,5}$}
%
\address{$^1$ SISSA, Via Bonomea 265, I-34136 Trieste, Italy}
\address{$^2$ CNR-IOM Democritos National Simulation Center, Via Bonomea 265, I-34136 Trieste, Italy}
\address{$^3$ Department of Physics, Bar-Ilan University (RA), Ramat Gan 52900, Israel}
\address{$^4$ Department of Physics, Indian Institute of Technology Kanpur, Kanpur 208016, India}
\address{$^5$ International Centre for Theoretical Physics (ICTP), P.O.Box 586, I-34014 Trieste, Italy}

\eads{\mailto{angelo.russomanno@tiscali.it}, \mailto{santoro@sissa.it} }

\begin{abstract}
We study the work statistics of a periodically-driven integrable closed quantum system, addressing in particular the role
played by the presence of a quantum critical point. 
Taking the example of a one-dimensional transverse Ising model in the presence of a spatially homogeneous but periodically time-varying
transverse field of frequency $\omega_0$, we arrive at the characteristic cumulant generating function $G(u)$, 
which is then used to calculate the work distribution function $P(W)$.  
By applying the Floquet theory we show that, in the infinite time limit, $P(W)$ converges, at zero temperature, 
towards an asymptotic steady state value whose small-$W$ behaviour depends only on the properties of the 
small-wave-vector modes and on a few important ingredients:  
the time-averaged value of the transverse field, $h_0$, the initial transverse field, $h_{\rm i}$, and the equilibrium 
quantum critical point $h_c$, which we find to generate a sequence of non-equilibrium critical points 
$h_{*l}=h_c+l\omega_0/2$, with $l$ integer.
When $h_{\rm i}\neq h_c$, we find a ``universal'' edge singularity in $P(W)$ 
at a threshold value of $W_{\rm th}=2|h_{\rm i}-h_c|$ which is entirely determined by $h_{\rm i}$. 
The form of that singularity --- Dirac delta derivative or square root --- depends on $h_0$ being or not at a non-equilibrium critical point $h_{*l}$. 
On the contrary, when $h_{\rm i}=h_c$, $G(u)$ decays as a power-law for large $u$, leading to different types of edge singularity 
at $W_{\rm th}=0$. 
Generalizing our calculations to the finite temperature case, the irreversible entropy generated by the periodic driving is also shown 
to reach a steady state value in the infinite time limit.
\end{abstract}

\pacs{75.10.Pq, 05.30.Rt, 03.65.-w}

\maketitle
\tableofcontents

\section{Introduction} 

In recent years, there have been many theoretical studies aimed at understanding the non-equilibrium dynamics of 
closed quantum systems\cite{Polkovnikov_RMP11,dutta:book}, inspired by a series of experiments on cold atomic gases,  
which are nearly isolated systems with long phase-coherence times, allowing for the study of a coherent quantum dynamics over long time scales
\cite{greiner02,kinoshita06,Jaksch_AP05,braun14}. These experiments have paved the way to addressing many fundamental questions, such as 
the role of integrability in thermalization following a quench \cite{Polkovnikov_RMP11} or the universal scaling of the defects generated 
when a system is driven across a quantum critical point \cite{dutta:book, dziarmaga10}. 
Usually, a quantum system is driven out of equilibrium by a slow ramping (annealing) or by a sudden quench of a parameter of the Hamiltonian 
(for example, the transverse magnetic field in the quantum Ising model discussed in this paper); 
the subsequent dynamical response of the system is encoded in several quantities, e.g.,  the Loschmidt echo \cite{quan06,pollmann10,sharma12,mukherjee12,nag12}, dynamical correlation functions \cite{calabrese06}, 
the growth of entanglement entropy \cite{Calabrese_JSTAT05}, or the change in diagonal entropy \cite{polkovnikov11}. 
In parallel, in the context of the Jarzynski equality \cite{jarzynski97} and non-equilibrium fluctuation relations \cite{Campisi_RMP11}, 
the question of the emergence of thermodynamical laws from a finite quantum system driven 
out of equilibrium and the generation of irreversible entropy have been addressed in several recent works \cite{Deffner10,Dorner_PRL_12}.

One of the ways to characterize the dynamics of an out-of-equilibrium quantum system  
is to explore the statistics of the performed work~\cite{Silva_PRL08, Gambassi_11:preprint, Gambassi12}. 
Given the non-equilibrium nature of the driving protocol, the work ($W$) is a stochastic variable and hence described by a probability 
distribution $P(W)$. 
Recently, from the analysis of the work statistics in systems quenched across a critical point, 
it has been shown that non-analyiticities in $P(W=0)$ at some critical times can mark the existence of a 
sequence of dynamical phase transitions in real time~\cite{heyl13}.
Furthermore, the knowledge of $P(W)$ enables us to obtain information about some universal features, by connecting it to the critical 
Casimir effect, for sudden quenches ending near the critical point \cite{Gambassi_11:preprint}; 
in particular, there exists a power-law edge singularity in $P(W)$, for small $W$, which is characterized by an exponent that is independent of the choice of protocol, but rather depends just on the initial and final values of the parameter being quenched \cite{Pietro_PRE}. 
It is usually convenient to define and work with the characteristic function $G(u)$, obtained by Fourier transforming $P(W)$. 
The characteristic function $G(u)$ has been shown to be closely related to the Loschmidt echo of a quenched quantum system, both at zero \cite{quan06,Silva_PRL08} and finite temperature \cite{Zanardi_PRA_07}. 

In this work we focus on a periodically driven integrable closed quantum system, namely the transverse Ising chain 
\cite{Mucco_JSM09, Arnab_PRB10}, with a spatially homogeneous but time-periodic transverse field 
$h(t)=h(t+2\pi/\omega_0)$, 
and calculate $P(W)$ stroboscopically at the end of $n$ complete periods $\tau=2\pi/\omega_0$. 
It has been shown in the literature \cite{Russomanno_PRL12,Russomanno_JSTAT13} that the system reaches a {\em periodic steady state} 
in the limit $n \to \infty$: the residual energy (which is in fact the first moment of $P(W)$) 
--- and indeed essentially any local observable --- reaches a stationary value, when observed at times $t_n=n\tau$, 
in the thermodynamic limit. We have observed a similar relaxation to a periodic steady condition 
also for a genuinely quantum non-local object, the so-called dynamical fidelity~\cite{Sharma_EPL14}.
In the present work we will investigate what happens to $G(u)$, and hence to $P(W)$, under such periodic driving in the asymptotic limit 
$n\to \infty$, and address the question of the universal behaviour emerging in the small-$W$ region.

Working within the framework of the Floquet theory \cite{Shirley_PR65,Grifoni_PR98}, and assuming that the initial state
is a Gibbs state at temperature $T$, we provide an analytical form of the characteristic function $G(u)$ in terms of Floquet quasi-energies 
and corresponding overlaps between the initial state and the Floquet eigenstates. 
$G(u)$ is then used to arrive at an exact expression of $P(W)$ at zero temperature: 
we demonstrate that indeed $P(W)$ also tends to ``synchronize'' with the periodic driving in the limit $n \to \infty$, 
converging to a well-defined asymptotic work distribution function $P_\infty(W)$,
whose small-$W$ behaviour depends only on the properties of the small-wavevector modes, ultimately controlled by 
the time-averaged value of the transverse field $h_0=\tau^{-1}\int_0^{\tau} h(t) \ud t$,
and its initial value $h_{\rm i}=h(t=0)$. 
%
%
All universal features of the work distribution are encoded into a singularity of $P(W)$ for small $W$. 
The position $W_{\rm th}=2|h_{\rm i}-h_c|$ of the edge of that singularity depends only on the distance 
of $h_{\rm i}$ from the equilibrium critical point $h_c$, while its detailed form 
--- a Dirac delta derivative or a square root singularity ---
depends on $h_0$ being or not at a {\em non-equilibrium critical point} $h_{*l}$
--- determined by $h_{*l}=h_c+l\omega_0/2$ with $l$ integer --- where the Floquet spectrum turns out to be gapless.
When $h_{\rm i}= h_c$, the generating function $G(u)$ has a power law decay $1/(iu)^p$, 
which results into a step singularity in $P_\infty(W)$ at $W_{\rm th}=0$ for $p=1$, or a mild $W^{p-1}$ increase for larger integer $p$. 
The non-equilibrium phase transitions we find are reminescent of those discussed in Ref.~\cite{Batisdas_PRA12}, 
but are different in many aspects, notably in residing at low frequency, as we will discuss.

An important aspect of these results is that the small-$W$ properties of $P_\infty(W)$ are
determined (as we are going to show in detail in the paper) exclusively by the small-$k$ modes: the aspects of the dynamics
independent of the details of the driving protocol rely only on the large wavelenght modes which encode the universal
properties of the ground state in the static system~\cite{Sachdev:book,Wilson_PR74}. Remarkably, a similar fact can
be observed in two coupled Luttinger liquids undergoing a quantum quench~\cite{Emanuele_PRL13}: 
if the quenched operator is relevant in the renormalization group sense (and then affecting mainly 
the long wavelength modes) then the phase coherence evolves with a universal scaling function.

Finally we move to the finite temperature case: starting from an initial mixed state at finite temperature, we will show
that the irreversible entropy generated during the periodic driving tends
to ``synchronize'' with the driving for $n \to \infty$. 
It saturates to a steady state value for large $\omega_0$, displaying a sequence of dips and
peaks for small and intermediate values of $\omega_0$, respectively. 
We note in passing a recent study on the work statistics of a periodically driven system which has explored the universal properties 
of the rate function, that is found to satisfy a lower bound and has a zero when $W$ matches the residual energy \cite{dutta14}.

The structure of the paper is as follows. In Sec.~\ref{Gu_intro:sec} we summarize the basic definitions and properties of
$P(W)$ and $G(u)$. Next, in Sec.~\ref{Ising_intro:sec} we focus on the case of a quantum Ising
chain undergoing a uniform generic periodic driving, showing in Sec.~\ref{Sec4} that the stroboscopic $G(u)$ and $P(W)$ converge 
to an asymptotic value (see~\ref{App:convergence} for mathematical details). In Sec.~\ref{Sec5} we discuss the behaviour of the asymptotic
$P(W)$ in the small-$W$ regime, showing that it is independent of the details of the driving protocol; we substantiate our analysis
with numerical results obtained in the case of a sinusoidal driving. In Sec.~\ref{Sec6} we discuss briefly the case with finite temperature $T$
in the context of irreversible entropy generation, and in Sec.~\ref{Sec7} we draw our conclusions together with perspectives of future work.
Technical details of our derivations are contained in three appendices.

\section{The work distribution $P(W)$ and its characteristic function $G(u)$} \label{Gu_intro:sec}
We present here, for the readers' convenience, some basic facts about the
work distribution function, following Refs.~\cite{Campisi_RMP11,Talkner_PRE07,Talkner_PRE08}.
Suppose that a closed quantum system undergoes a time-dependent driving
such that its Hamiltonian is $\op{H}(t)$, while the system was at the initial time
$t_0=0$ in a given (possibly mixed) state $\op{\rho}(0)$.
%
If $p_n(0)$ is the probability that the system has energy $E_n(0)$ at the initial time,
and $P(m \tf |n 0)$ the conditional probability that the system is observed
to have energy $E_m(\tf)$ at some later time $\tf$, then the work distribution
function is defined as:
\begin{equation}
P_{\tf}(W) = \sum_{n,m} \delta(W-E_m(\tf)+E_n(0)) \; P(m \tf | n 0) \; p_n(0) \;.
\label{eq_work_dis}
\end{equation}
%
%
%
To deal with the Dirac's deltas appearing in the definition of $P_{\tf}(W)$ it is convenient to study the Fourier transform 
of $P_{\tf}(W)$, arriving at the characteristic function:
\begin{equation}
G_{\tf}(u) = \int_{-\infty}^{+\infty} \!\! dW \nep^{iuW} P_{\tf}(W) \;.
\end{equation}
With simple manipulations, and introducing the unitary evolution operator $\op{U}(\tf,0)$ for the closed quantum system, 
it results that: 
\begin{equation} \label{G_final:eqn}
G_{\tf}(u) = \Tr \Big( \op{U}^{\dagger}(\tf,0) \nep^{iu \op{H}(\tf)} \op{U}(\tf,0) \nep^{-iu \op{H}(0)} \op{\rho}(0)  \Big)
= \Tr \Big(  \nep^{iu \op{H}_H(\tf)} \nep^{-iu \op{H}(0)} \op{\rho}(0)  \Big) \;,
\end{equation}
where $\op{H}_H(\tf)= \op{U}^{\dagger}(\tf,0) \op{H}(\tf) \op{U}(\tf,0)$ is the final Hamiltonian in Heisenberg representation,
%
%
and we have assumed that the initial state is such that $[\op{\rho}(0),\op{H}(0)]=0$ (a Gibbs or micro-canonical state would do that).
%
%
We note, in passing, that the quantum Jarzynski equality~\cite{jarzynski97,Dorner_PRL_12} follows immediately from the previous expression 
by taking $u=i\beta$ and assuming an initial Gibbs state $\op{\rho}(0)=\nep^{-\beta (\op{H}(0)-F_0)}$:
\begin{equation}
G_{\tf}(i\beta) = \int_{-\infty}^{+\infty} dW \nep^{-\beta W} P_{\tf}(W) = \nep^{-\beta (F_{\rm f}-F_0)} \;. 
\end{equation}

\section{The uniform periodically driven quantum Ising chain} \label{Ising_intro:sec}
Let us now specialize our discussion to the quantum Ising chain in a uniform time-periodic transverse field 
(although some progress might be done in the general non-homogeneous case).  
The Hamiltonian we consider is \cite{Sachdev:book}:
\begin{equation}  \label{hamil}
   \op{H}(t) = -\frac{J}{2} \sum_{j=1}^{L} \left[ \ops{\sigma}_j^z \ops{\sigma}_{j+1}^z + h(t) \ops{\sigma}_j^x \right] \;,
\end{equation}
where $h(t)=h(t+\tau)$ is a generic uniform-in-space transverse field which is periodically driven with frequency $\omega_0 = 2\pi/\tau$
around the average value
\begin{equation} \label{average_h:eqn}
  h_0 = \frac{1}{\tau}\int_0^\tau h(t)\,\ud t\;.
\end{equation}
%
The Hamiltonian can always be recast in the form of a quadratic fermionic model, thanks to the 
Jordan-Wigner transformation \cite{Lieb_AP61}.
Upon Fourier transforming in space to the relevant Jordan-Wigner fermions, we get:
\begin{equation} \label{hamil1}
\op{H}^{+}(t)=\sum_k^{\rm ABC} \op{H}_k(t) = 
  \sum_{k}^{\rm ABC} \left[ \begin{array}{cc} \opcdag{k} & \opc{-k} \end{array} \right] \Big[ {\mathbb H}_k(t) \Big] 
  \left[ \begin{array}{l} \opc{k} \\ \opcdag{-k} \end{array} \right] \;,
\end{equation}
where the $2\times 2$ matrix ${\mathbb H}_k(t)$ has the form:
\begin{equation} \label{formaggioska:eqn}
 {\mathbb H}_k(t)  = \left[ \begin{array}{cc}
			           	\epsilon_k(t) 	&-i\Delta_k\\
					i\Delta_k		&-\epsilon_k(t) \end{array} \right] \;,
\end{equation}
with $\epsilon_k(t) = h(t)-\cos k$ and $\Delta_k = \sin k$, having set $J=1$.
Notice that the previous Hamiltonian is really only a part of the total $\op{H}=\op{H}^{+}+\op{H}^{-}$, 
i.e., the part living in the subspace with even fermion-parity, for which anti-periodic boundary conditions (ABC) apply,
and $k=(2n-1)\pi/L$ with $n=1 \cdots L/2$ \cite{Lieb_AP61}. 
%
%
This is certainly enough for describing the ground state and the dynamics starting from the ground state.
At finite temperature, the contributions due to the extra odd-fermion-parity term,  $\op{H}^{-}$, corresponding
to periodic boundary conditions $k$-values, are automatically accounted for when we transform the sum over $k$ 
into an integral over the whole Brillouin zone $k\in[0,\pi]$. 

%
%
%
%
We assume that the system is, at time $t=0$, in the Gibbs ensemble at temperature $T$ for the Hamiltonian $\Ham(0)$. 
Following a standard procedure \cite{Russomanno_PRL12,Russomanno_JSTAT13,Sharma_EPL14},
the initial problem is diagonalized by introducing new fermionic operators
$\opgammadag{k}=u_k \opcdag{k} + v_k \opc{-k}$, in terms of which the BCS-ground state 
for each $k$ reads $|\psi_{k}^{\rm gs}\rangle= [ u_k + v_k \opcdag{k} \opcdag{-k} ] |0\rangle$,
with energy $-E_k=-\sqrt{\epsilon_k^2(0)+\Delta_k^2}$, and the excited state is
$|\psi_{k}^{\rm ex}\rangle= \opgammadag{k}\opgammadag{-k} |\psi_{{\rm gs}, k}\rangle 
= [ v_k + u_k \opcdag{k} \opcdag{-k} ] |0\rangle$, with energy $+E_k$.
If we are interested in calculating $G_{\tf=n\tau}(u)$ at integer multiples of the period $\tau=2\pi/\omega_0$,
it is sufficient to construct the Floquet modes $|\phi_{k}^{\pm}(0)\rangle$ with corresponding Floquet
quasi-energies $\pm \mu_k$; this must be done, in general, by a numerical integration of the
Schr\"odinger equation, for each $k$, over a single period \cite{Russomanno_PRL12, Sharma_EPL14}. 
The outcome of that calculation provides the relevant overlaps 
$r_{k}^{\pm}=\langle \phi_{k}^{\pm}(0) | \psi_{k}^{\rm gs}\rangle$, with $|r_{k}^+|^2+|r_{k}^-|^2=1$, and, by unitarity, 
$\langle \phi_{k}^{\pm}(0) | \psi_{k}^{\rm ex}\rangle=\mp (r_{k}^{\mp})^*$.
With these basic ingredients, the derivation of a general expression of $G_{n\tau}(u)$ for the uniformly driven Ising 
chain follows essentially the steps outlined in Ref.~\cite{Sharma_EPL14}, generalized to an arbitrary finite temperature. 
The final result can be cast in the form:
%
%
\begin{equation} \label{logG_Ising:eqn}
\ln G_{n\tau}(u) = \sum_{k>0}^{\rm ABC} \ln 
\Big\{ 1- \frac{2q_k}{1+q_k} \sin^2{(\mu_k n\tau)} \Big[ (1-\nep^{2iu E_{k}}) (1-f_k) + 
                                             (1-\nep^{-2iu E_{k}}) f_k \Big] \Big\} \hspace{0mm} \;,
\end{equation}
where $f_k=1/(\nep^{2\beta E_k} +1)$ denotes the Fermi occupation function 
(observe that creating an excitation costs here an energy $2E_k$) and
\begin{equation} \label{qk_def:eqn}
q_k \equiv \frac{2\left|r_k^+\right|^2\left|r_k^-\right|^2}{\left|r_k^+\right|^4+\left|r_k^-\right|^4} =
\frac{2\left|r_k^+\right|^2\left|r_k^-\right|^2}{1- 2\left|r_k^+\right|^2\left|r_k^-\right|^2} \;.
\end{equation}
Had we chosen to work with the Laplace transform of the $P(W)$, rather than with the Fourier transform $G(u)$, 
we would have obtained a similar expression with the formal replacement $u\to is$. 
For $T=0$ we would get 
\begin{equation} \label{logG_IsingT0:eqn}
\ln G_{n\tau}^{T=0}(is) = \sum_{k>0}^{\rm ABC} \ln 
\Big\{ 1- \frac{2q_k}{1+q_k} \sin^2{(\mu_k n\tau)} (1-\nep^{-2s E_{k}}) \Big\} \;,
\end{equation}
which actually shows that the expression for $G(is)$ is better behaved at $T=0$, 
at the expense of having to perform an inverse Laplace transform to recover $P(W)$. 
In the thermodynamic limit, transforming the sum over $k$ into an integral, we can finally write:
\begin{equation} \label{logG_final_IsingT0:eqn}
\frac{\ln G_{n\tau}^{T=0}(is)}{L} \stackrel{\scriptscriptstyle L\to \infty} {\longrightarrow}
\int_{0}^{\pi} \! \frac{\ud k}{2\pi} \; 
\ln \Big\{ 1- \frac{2q_k}{1+q_k} \sin^2{(\mu_k n\tau)} (1-\nep^{-2s E_{k}}) \Big\} 
\;.
\end{equation}
This object (see also Eq.~\eqref{cumulant:eqn}) is the so-called {\em cumulant generating function} and coincides
\footnote{The relevant quantity in Ref.~\cite{Pietro_PRE} translates as follows in terms of our quantities: 
\[ |y_k(n\tau)|^2 \equiv \frac{2q_k \sin^2{(\mu_k n\tau)} }{[1+q_k\cos{(2\mu_k n\tau )}]} . \]
We also mention that the analysis of Ref.~\cite{Pietro_PRE} identifies the large-$s$ limit of 
$\ln G_{n\tau}^{T=0}(is)/L$ with a ``surface'' free-energy contribution 
\[ -2f_{\rm surf}= \int_0^\pi \! \frac{\ud k}{2\pi} \; \ln \left[1+\left|y_k(n\tau) \right|^2\right] \;, \]
which coincides with our $g_n$. 
}
with the expression of Smacchia {\em et al.} \cite{Pietro_PRE}.
%
%
We observe also that, in the limit $s\to \infty$, we recover the expression for $g_n$, the logarithm 
of the dynamical fidelity $\mathcal{F}(n\tau)=\left| \langle \Psi_0 | \Psi(n\tau)\rangle \right|^2=\nep^{Lg_n}$ 
discussed in Ref.~\cite{Sharma_EPL14} (see Eq.~(8) there).

\section{Steady state of the work probability distribution $P_{n\tau}(W)$ for $n\to \infty$} \label{Sec4}
\begin{figure}
\begin{center}
\includegraphics[width=9cm]{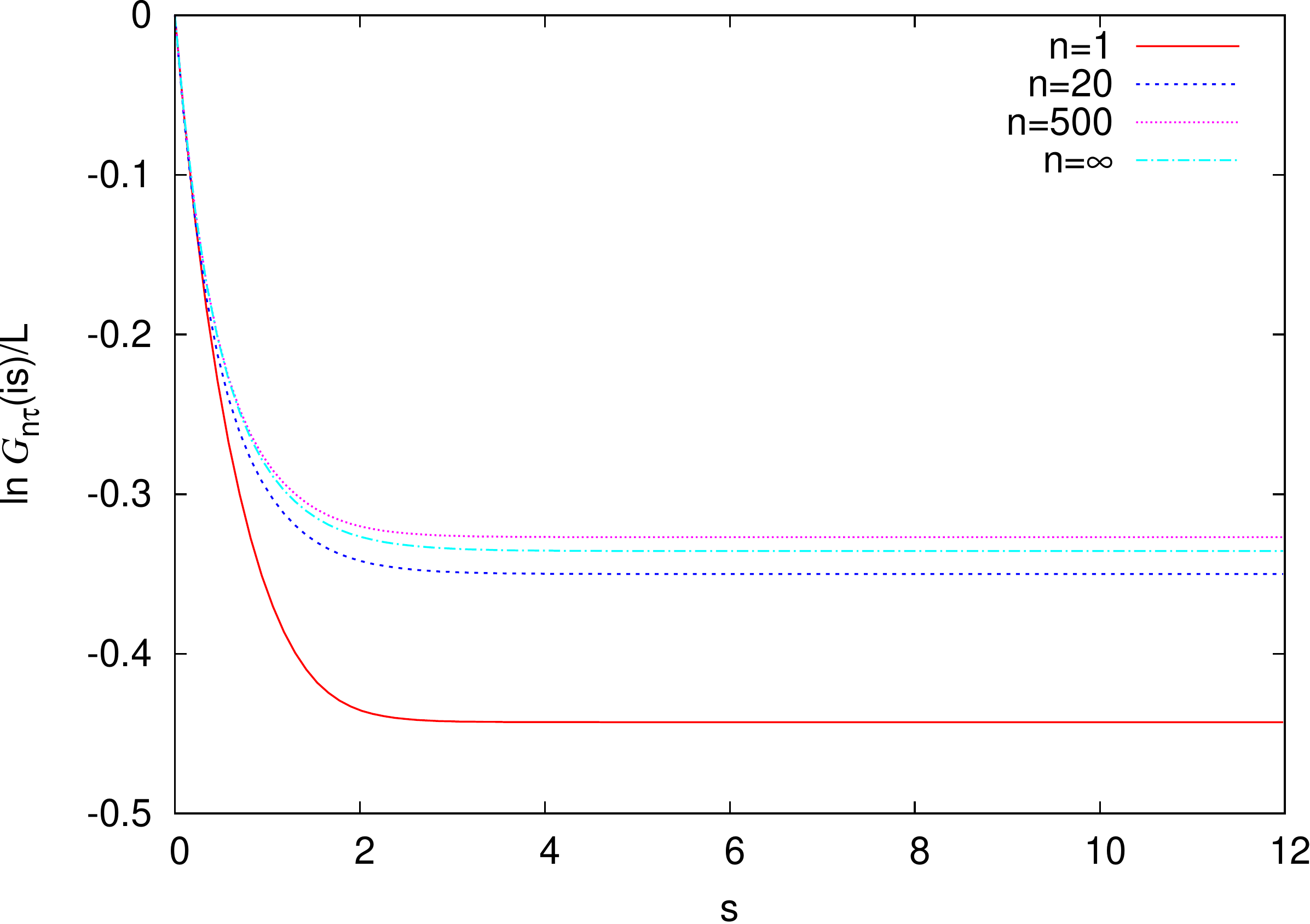}
\includegraphics[width=9cm]{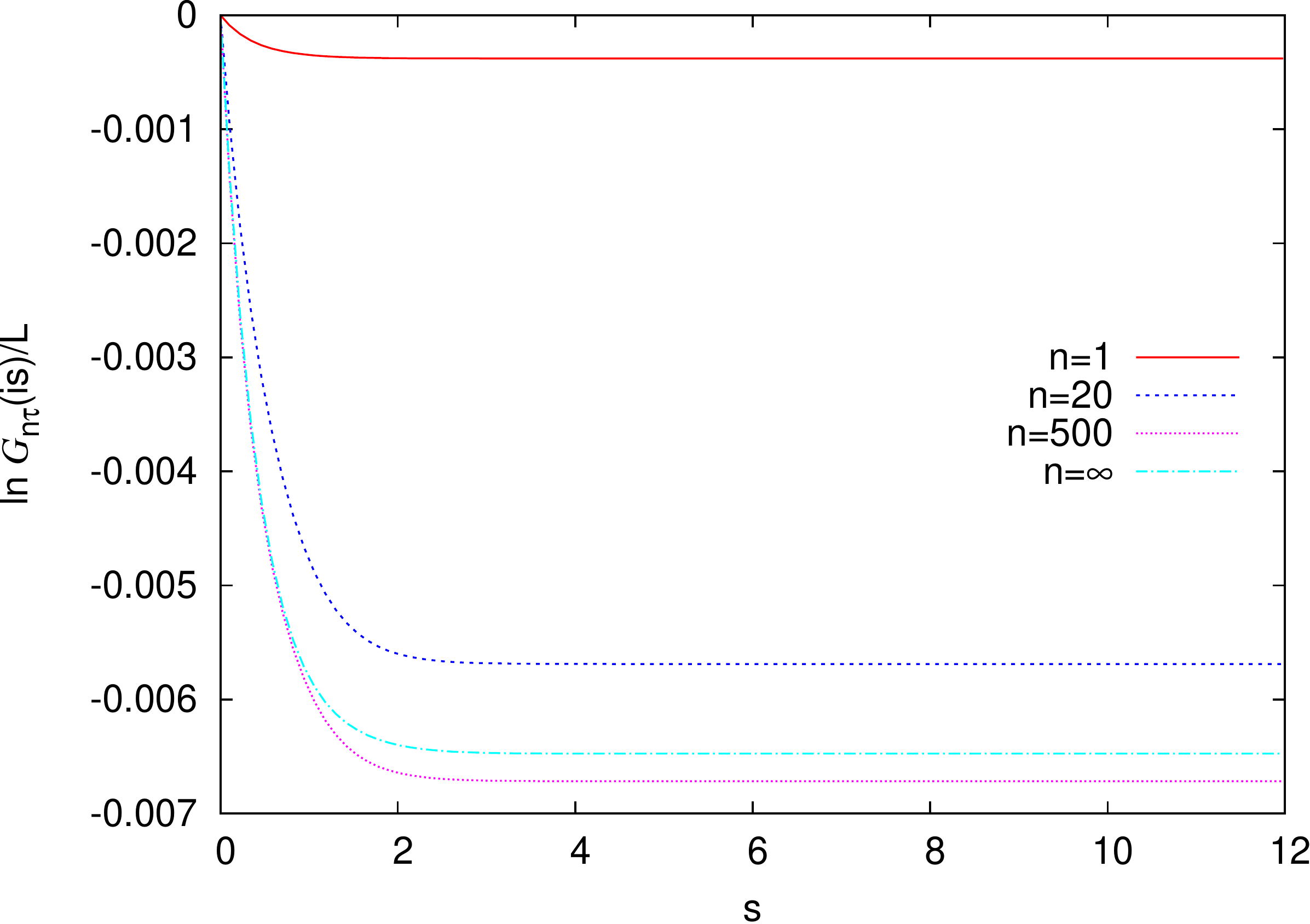}
\end{center}
\caption{Plot of $\frac{\ln{G}_{n\tau}(is)}{L}$ versus $s$ for different $n$, compared to $\frac{\ln{G}_{\infty}(is)}{L}$, for a 
Ising model with periodic driving $h(t)=h_0+A\cos{(\omega_0 t +\phi_0)}$; here $h_0=h_c=1$, $A=1$,  
and two different values of the frequency are shown, $\omega_0=2$ (top) and 
$\omega_0=0.3623$ (bottom), corresponding to a situation in which $J_0(2A/\omega_0)\approx 0$ and
there is coherent destruction of tunnelling (see discussion in the text). Numerical data are for $L=1000$. 
}
\label{G_conv:fig}
\end{figure}

The question we are now going to address is if the probability distribution of the work $P_{\tf}(W)$ tends to ``synchronize''
with the periodic driving in the asymptotic limit $\tf \to \infty$. 
When viewing at the system stroboscopically at the times $t_n=n\tau$ (integer multiples of the period $\tau$), what
we want to understand is if $P_{n\tau}(W)$ converges towards a well defined asymptotic work distribution for $n\to \infty$. 
The answer is positive, as we are now going to show.
We know from previous work \cite{Russomanno_PRL12} that the quantum average of the work 
(i.e., the first moment of $P_{n\tau}(W)$) indeed reaches a ``steady state'' for $n\to \infty$, in the thermodynamic limit. 
\footnote{
Strictly speaking, when $L\to \infty$ the $P(W)$ becomes narrower and narrower, on the scale of the average work $\langle W\rangle\sim L$, with fluctuations which scale as $\sqrt{L}$.
Nevertheless, when $L$ is large but not $\infty$ and the approximation of having set $L\to \infty$ 
in transforming the sum into an integral holds only until a certain finite time $t^*\sim L$, 
the question we are asking is meaningful, provided the ``steady state'' is effectively reached before $t^*$.
}
Moreover, as we have recently shown in Ref.~\cite{Sharma_EPL14}, the $\delta(W)$ Dirac-delta part of the distribution $P_{\infty}(W)$
(see, for instance, Eq.~\eqref{Pinf} below)
{\it alias} the large-$s$ limit $g_n=\lim_{s\to\infty} \ln G^{T=0}_{n\tau}(is)/L$  --- which corresponds to the dynamical fidelity --- 
also reaches a well defined ``steady state'' for $n\to \infty$: $\lim_{n\to\infty} g_n = g_{\infty}$.
Our claim now is that {\it all the cumulants} of $P_{n\tau}(W)$ reach such a ``steady state'', and therefore so does the whole probability distribution. 
To see this, it is enough to show that the whole cumulant generating function, Eq.~\eqref{logG_final_IsingT0:eqn},
reaches a steady state. 
With an argument which generalizes that of Ref.~\cite{Sharma_EPL14}, whose details are given in~\ref{App:convergence}, 
one can show that when $n\to\infty$ this quantity tends to the stationary value
\begin{equation} \label{logG-asymp}
\frac{\ln G_{\infty}^{T=0}(is)}{L} \equiv \lim_{n\to\infty} \frac{\ln G_{n\tau}^{T=0}(is)}{L} = 
2 \int_0^\pi\! \frac{\ud k}{2\pi} \; \ln \left[\frac{1+\sqrt{1-\xi_k(s)}}{2} \right] \;,
\end{equation}
where 
\begin{equation} \label{xi_k:def}
\xi_k(s)\equiv 4\left|r_k^+\right|^2\left|r_k^-\right|^2\left(1-\nep^{-2s E_{k}} \right) \;.
\end{equation}
The numerical results shown in Fig.~\ref{G_conv:fig} perfectly confirm this analytic prediction: there is a transient 
--- whose details depend on the parameters, for instance on the average field $h_0$ --- which then leads to an asymptotic result
for $n\to \infty$ given by the simple analytic expression in Eq.~\ref{logG-asymp}.
Fig.~\ref{G_conv:fig} also illustrates (bottom panel) a case in which the frequency $\omega_0$ is such that
$J_0(2A/\omega_0)\approx 0$, where $J_0(x)$ is the $0$th-order Bessel function, and there is 
{\em coherent destruction of tunnelling} \cite{Grossmann_PRL91,Grifoni_PR98}: 
observe that $\frac{\ln{G}_{n\tau}(is)}{L}$ has a very small magnitude for this value of $\omega_0$.

The cumulants of the asymptotic work distribution are obtained as:
\begin{equation} \label{cumulant:eqn}
  K_m=(-1)^m \frac{\ud^m}{\ud s^m} \left. \ln G_{\infty}^{T=0}(is) \right|_{s=0} \;.
\end{equation}
The first cumulant (which is the quantum average of the work performed) is given by:
\begin{equation} \label{Work_infty_T=0:eqn}
  K_1=\mean{W}_\infty = 4 L \int_0^\pi\! \frac{\ud k}{2\pi}\; \left|r_k^+\right|^2\left|r_k^-\right|^2 E_{k} \;.
\end{equation}
This coincides with the result we would obtain by evaluating
\[ \lim_{n\to \infty} \int_0^\pi \frac{\ud k}{2\pi} \left[\bra{\psi_k(n\tau)}\hat{H}_k(0)\ket{\psi_k(n\tau)}-E_{k}^{\rm gs}(0)\right] \;, \]
which is quite easy to calculate directly.
The second cumulant is the variance of the work distribution and is given by:
\begin{equation}
  K_2 = \sigma_\infty^2 = L\; \int_0^\pi\! \frac{\ud k}{2\pi}\; \left[4\left|r_k^+\right|^2\left|r_k^-\right|^2
    \left(1+3\left|r_k^+\right|^2\left|r_k^-\right|^2\right) E_{k}^2 \right] \;.
\end{equation}
We notice that the $P(W)$ tends to become narrower and narrower in the thermodynamic limit, as expected, because
$\sigma_\infty/\mean{W}_\infty \propto 1/\sqrt{L}$.
%
%

\section{Universal edge singularity at small $W$ in $P_{\infty}(W)$} \label{Sec5}
\begin{figure}
\begin{center}
\includegraphics[width=9cm]{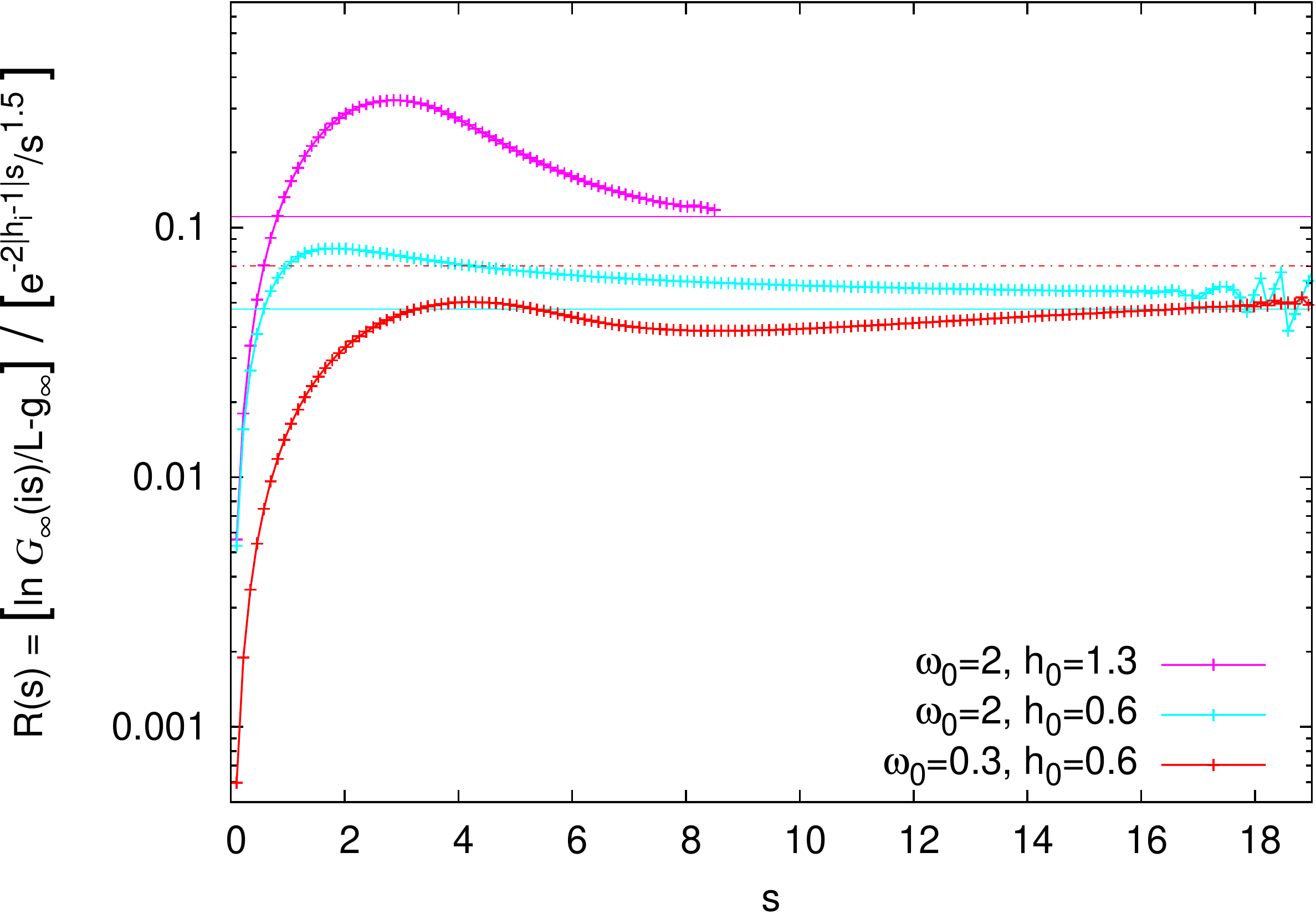}
\includegraphics[width=9cm]{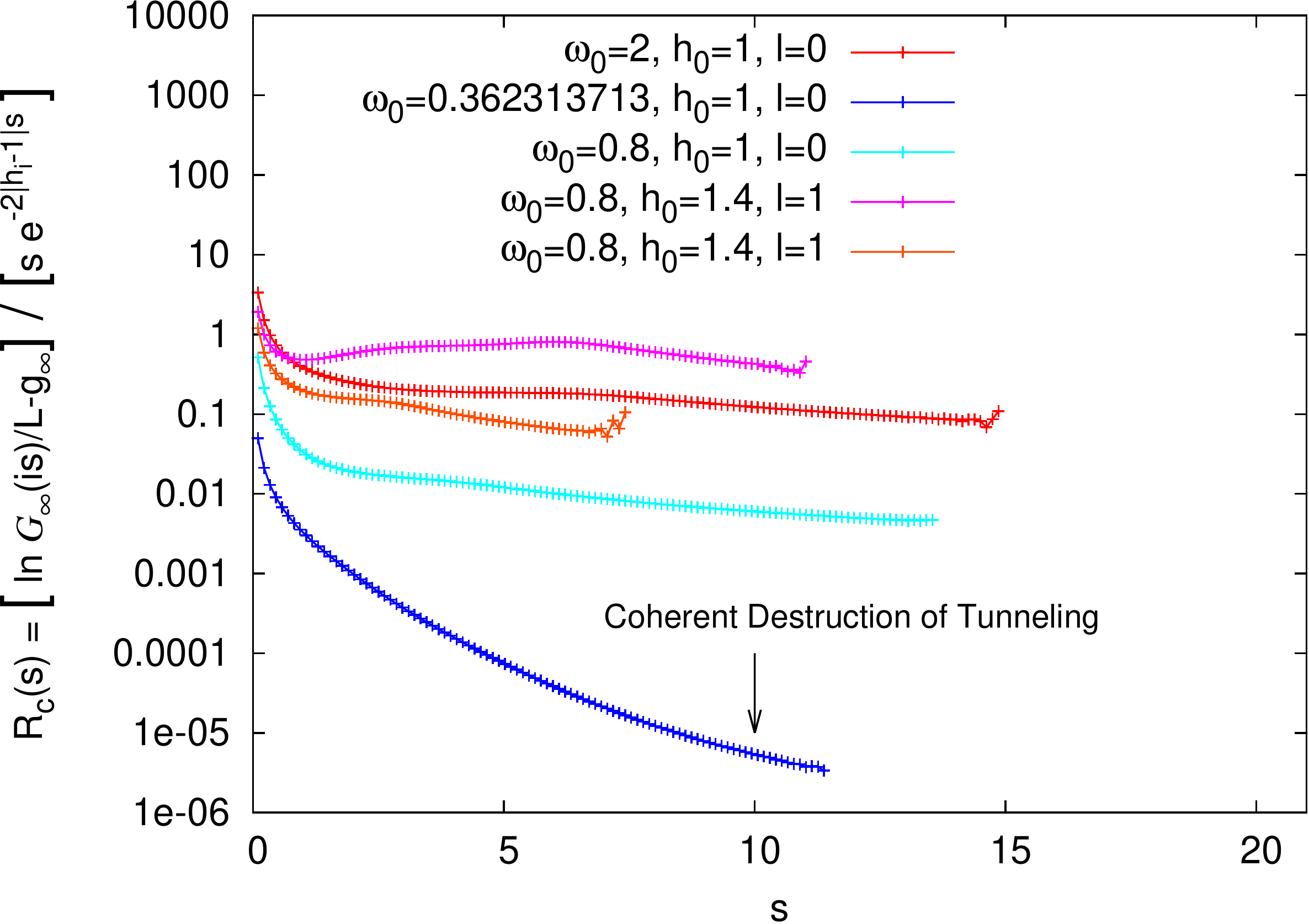}
%
\end{center}
\caption{Asymptotic cumulant generating function for a periodically driven uniform quantum Ising chain.
In these figures we assume a driving $h(t) = h_0 + A\cos(\omega_0 t+\phi_0)$, with $A=1$ and $\phi_0=0$.
(Top) Driving with $h_0\neq h_c=1$ and $h_{\rm i}\neq h_c=1$, plot of 
$R(s)=\left[\frac{\ln{G}_{\infty}(is)}{L}-g_\infty\right]/\left[\nep^{-2|h_{\rm i}-1|s}/s^{3/2}\right]$ versus $s$.
For large $s$, we see a convergence towards a finite limit ($a$, as defined in Eq.~\eqref{a_def:eqn} and calculated
by numerically fitting the coefficient $\alpha^2$ appearing in $|r_k^+|^2=\alpha^2k^2/4$ for small $k$).
(Bottom) Driving with $h_{\rm i}\neq h_c=1$ but $h_0$ fulfilling the Floquet resonance condition 
Eq.~\eqref{reso_cond:eqn} (we take instances with resonances at $l=0$ and $l=1$).
Plot of $R_c(s)=\left[\frac{\ln{G}_{\infty}(is)}{L}-g_\infty\right]/\left[s\nep^{-2|h_{\rm i}-1|s}\right]$ versus $s$, 
showing the convergence towards a finite limit for large $s$, in agreement with Eq.~\eqref{Gus_spec}.
Notice that cases where there is coherent destruction of tunnelling fail to be described by such a formula. 
}
\label{asymptotica:fig}
\end{figure}

Inspired by the results of Ref.~\cite{Pietro_PRE}, we now discuss the behaviour of the asymptotic
work probability distribution at small values of $W$, especially in connection with aspects which are independent
of the details of the specific driving protocol. 
From a technical point of view, the small-$W$ behaviour of $P_{\infty}(W)$ is encoded in the large-$s$ behaviour of $G_{\infty}(is)$,
which we can evaluate by means of Eq.~\eqref{logG-asymp}.
We will show that, indeed, the important ingredients are: 
{\it i)} the value $h_{\rm i}$ of the initial transverse field $h(t=0)$, and 
{\it ii)} the value of the average field $h_0$  (and the frequency $\omega_0$), determining if
the Floquet spectrum shows a resonance, and is gapless, at $k=0$, or not. 
Indeed, the value of $h_{\rm i}$ determines the {\em position} $W_{\rm th}$ of the singularity in $P(W)$ which we observe, while
the {\em form} of this singularity is determined by the possibility that the Floquet spectrum has a resonance at $k=0$,
which is entirely determined by the time-averaged field $h_0$ (see Eq.~\eqref{average_h:eqn}) hitting  
a {\em non-equilibrium} quantum critical point $h_{*l}=h_c+l\omega_0/2$, with $l$ integer, i.e., 
\begin{equation} \label{reso_cond:eqn}
  2|h_0-h_c| = l\omega_0 \quad\textrm{for some positive integer}\; l \;,
\end{equation}
%
%
is obeyed. This conclusion holds for a generic periodic driving $h(t)$: details can be found in ~\ref{App:small_k}. 
The small-$W$ universal behaviour of $P_\infty(W)$ we describe below relies, in the end, only on the properties of the small-$k$ modes,
in particular on the small-$k$ behaviour of $|r_k^+|^2$ in Eq.~\eqref{xi_k:def}: 
we find that if the resonance condition in Eq.~\eqref{reso_cond:eqn} is fulfilled
~\footnote{As we show in~\ref{App:small_k} there
are situations in which Eq.~\eqref{h02-bis:eqn} is valid for $|r_k^-|^2$ and not for $|r_k^+|^2$. Nevertheless, all the results
are unchanged: since $|r_k^+|^2+|r_k^-|^2=1$, exchanging the roles of $|r_k^+|$ and $|r_k^-|$ has no effect,
as Eq.~\eqref{logG-asymp} is symmetric under such an exchange.
}, then
\begin{equation} \label{h01-bis:eqn}
  |r_k^+|^2=\frac{1}{2}-\frac{1}{2}\beta k+O(k^2) \;;
\end{equation}
otherwise
\begin{equation} \label{h02-bis:eqn}
  |r_k^+|^2= \frac{\alpha^2}{4} k^2+O(k^4)\;.
\end{equation}
The precise values of $\alpha$ and $\beta$ depend on the specific form of driving, but the functional form of $|r_k^+|^2$ 
and the Floquet spectrum being gapless at $k=0$ or not depend only on the fulfillment of Eq.~\eqref{reso_cond:eqn}.
Depending on $h_{\rm i}$ and $h_0$, we can distinguish essentially three different behaviours of $P_{\infty}(W)$
in the small-$W$ limit:

%
{\it Case a) $h_{\rm i}\neq h_c$ and non-resonant Floquet spectrum.} 
In this case we obtain (see \ref{App:low_energy} for the derivation) an approximate analytical formula  
for $ G_{\infty}(is)$, valid when $s\gg 1/|h_{\rm i}-h_c|$:
\begin{equation}  \label{Gus1}
  G_{\infty}(is) \simeq \nep^{Lg_{\infty}} \left( 1 +\frac{aL}{{s}^{3/2}} \nep^{-2s \left|h_{\rm i} - h_c\right|}\right) \;.
\end{equation}
The inverse Laplace transform predicts that the small-$W$ behaviour of $P_{\infty}(W)$ is given by
\footnote{
 Remarkably, the form of the singularity in Eq.~\eqref{Pinf} for the asymptotic work-distribution function
 is the same found in \cite{Pietro_PRE} in the case of a generic quench starting and ending into the same
 paramagnetic or ferromagnetic case. This is exactly what we are doing in this periodic driving protocol.} 
\begin{equation} \label{Pinf}
P_{\infty}(W) \simeq \nep^{Lg_{\infty}} \left( \delta(W) +\frac{2aL}{\sqrt{\pi}} \; \sqrt{W-2\left|h_{\rm i}-h_c\right| }  \; 
                             	\theta\left(W-2\left|h_{\rm i}-h_c\right|\right) \right) \;,
\end{equation}
which applies whenever $W \lesssim (2+\ln 2)|h_{\rm i}-h_c|$. 
In both expressions the constant $a$ 
is given by: 
\begin{equation} \label{a_def:eqn}
  a\equiv \frac{\alpha^2}{16\sqrt{\pi}}\left(\frac{|h_{\rm i}-h_c|}{h_{\rm i}}\right)^{3/2} \;,
\end{equation}
where $\alpha^2$ is such that $|r_k^+|^2\simeq\alpha^2 k^2/4$ for small $k$ (see Eq.~\eqref{h02-bis:eqn}).
Eq.~\eqref{Pinf} predicts an edge singularity in the asymptotic work distribution function at a
precise value of $W$ which is totally independent of the details of the periodic protocol (and even of the frequency) 
but depends only on the initial value $h_{\rm i}$ of the field.
The details of the protocol enter into the strength of the singularity (the coefficient $a$). 
We notice that the threshold $2\left|h_{\rm i}-h_c\right|$ is the energy which has to be provided to the system
to generate an excitation in the $k=0$ mode; moreover, also the form of the singularity is only determined, through the
constant $a$, by the small-$k$ Floquet modes, as detailed in~\ref{App:small_k}. 
So, the behaviour at small $W$ of the work distribution function is dominated by the modes of lowest energy: 
in retrospective, this is a very reasonable finding. 
We stress that the previous analytical expressions are approximations valid in a precise range of $s$ or $W$. 
The only condition for the validity of these approximate formulas, as detailed in~\ref{App:low_energy}, is that the driving field 
does not start from the critical point value, i.e., $h_{\rm i}\neq h_c$ and there are no resonances at $k=0$ in the Floquet spectrum. 
We show some instances of the validity of Eq.~\eqref{Gus1} in the upper panel of Fig.~\ref{asymptotica:fig},
where we numerically evaluate $\frac{\ln{G}_{\infty}(is)}{L}$ for $h(t) = h_0 + A\cos(\omega_0 t+\phi_0)$ and plot 
$R(s)\equiv\left[\frac{\ln{G}_{\infty}(is)}{L}-g_\infty\right]/\left[\nep^{-2|h_{\rm i}-1|s}/{s}^{3/2}\right]$ versus $s$ for
different values of $h_0\neq h_c$, $h_{\rm i}\neq h_c$ and $\omega_0$. 
We see that, for large $s$, $R(s)$ tends towards a constant, which Eq.~\eqref{Gus1} predicts to be $a$, Eq.~\eqref{a_def:eqn}, 
denoted by horizontal lines (the values of $\alpha^2$ were obtained by a numerical fitting of $|r_k^+|^2$ for small $k$, 
according to Eq.~\eqref{h02-bis:eqn}). 
Overall, we see that there is a good agreement with the asymptotic value of $R(s)$.
%
%

{\it Case b) $h_{\rm i}\neq h_c$ and resonant Floquet spectrum (Eq.~\eqref{reso_cond:eqn} fulfilled).} 
In this case the Gaussian analysis performed in ~\ref{App:low_energy} fails.
We find, nevertheless, an approximate form of ${G}_{\infty}(is)$ for large $s$ ($s\gg 1/|h_{\rm i}-h_c|$) as
\begin{equation}\label{Gus_spec}
  G_{\infty}(is) \simeq \nep^{Lg_{\infty}} \left( 1 + L\, a_c\, s\, \nep^{-2s\left|h_{\rm i}-h_c\right|} + \cdots \right) \;,
\end{equation}
%
as long as there is no coherent destruction of tunnelling, i.e., for a resonance of order $l$, we have that $J_l(2A/\omega_0)\neq 0$
($J_l$ being the Bessel function of the first kind of order $l$~\cite{Stegun}). 
The resulting $P(W)$, after inverse Laplace transform, has a very singular contribution  
\begin{equation} \label{Pinf-h0=1}
P_{\infty}(W) \simeq \nep^{Lg_{\infty}} \Big( \delta(W) 
+a_c\,L\, \delta'(W-2\left|h_{\rm i}-h_c\right|) \; {\theta\left(W-2\left|h_{\rm i}-h_c\right|\right)} +\cdots \Big) \;.
\end{equation}
The lower panel of Fig.~\ref{asymptotica:fig} shows plots of
$R_c(s)\equiv \left[\frac{\ln{G}_{\infty}(is)}{L}-g_\infty\right]/\left[s\nep^{-2|h_{\rm i}-1|s}\right]$ for different resonant cases
with $l=0$ and $l=1$.
%

\begin{figure}
\begin{center}
\includegraphics[width=9cm]{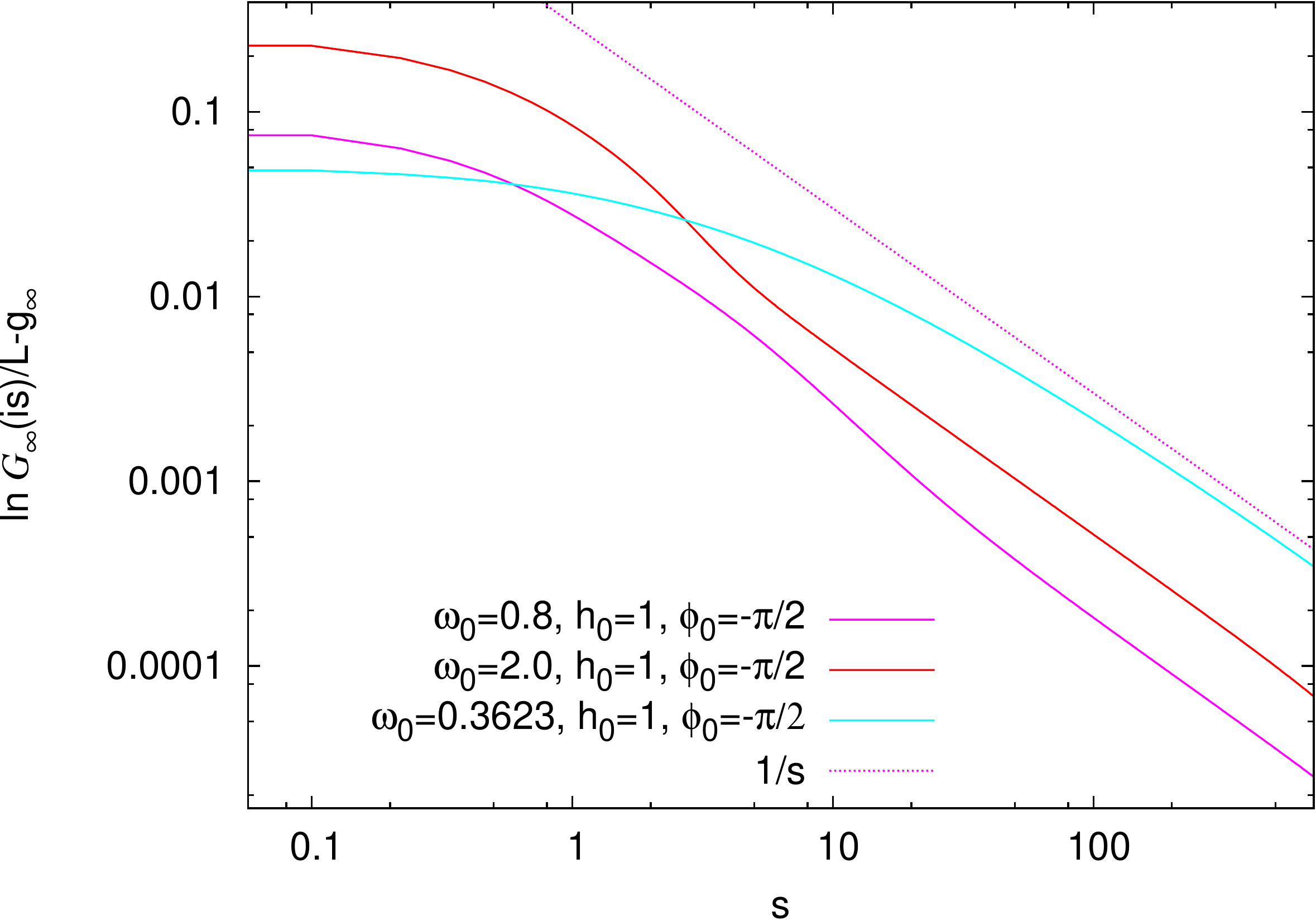}\\
\quad\\
\includegraphics[width=9cm]{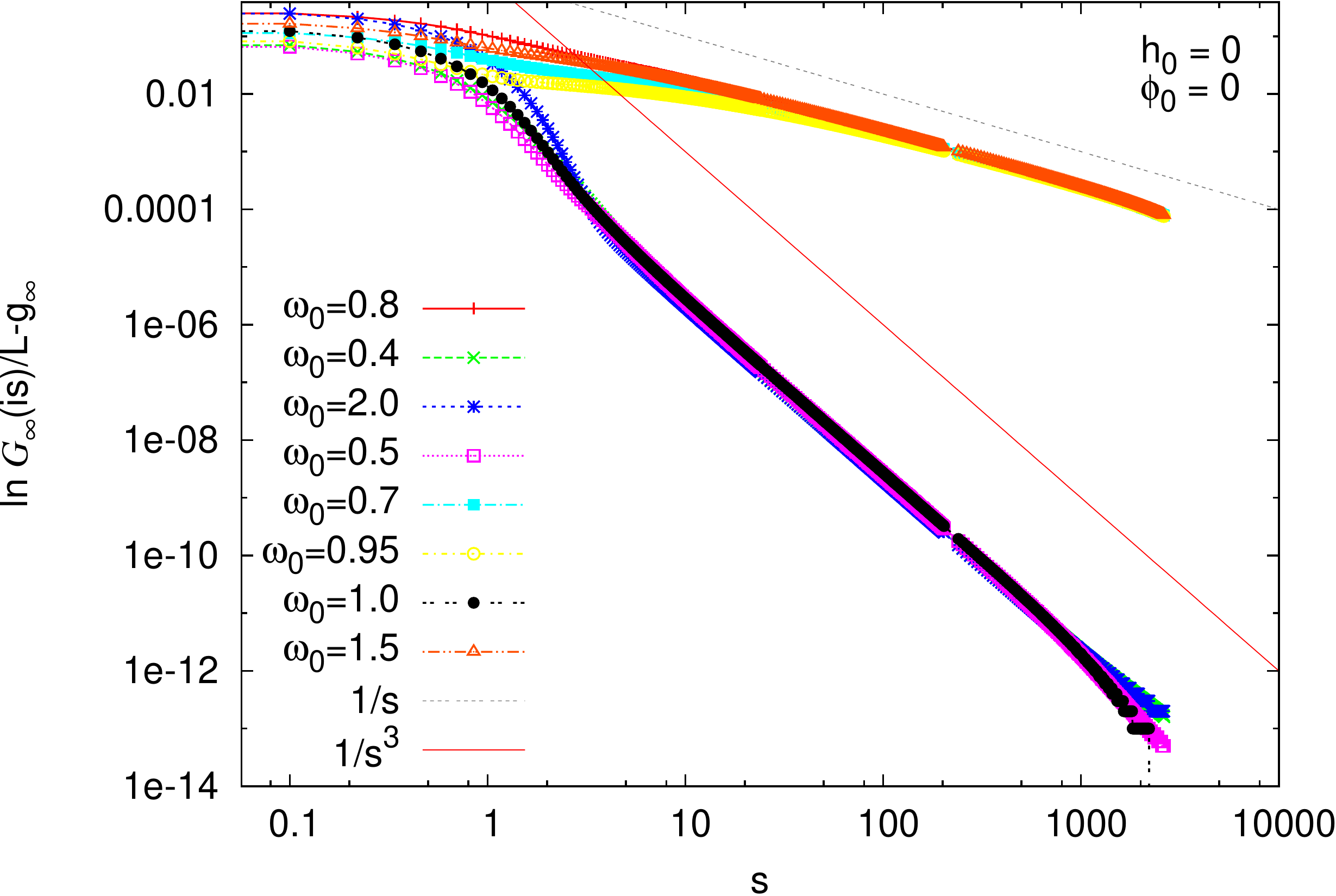}
\end{center}
\caption{Asymptotic cumulant generating function for a periodically driven uniform quantum Ising chain.
In these figures we assume a driving $h(t) = h_0 + A\cos(\omega_0 t+\phi_0)$, with $A=1$.
(Top) Instances of the critical case with $h_{\rm i}=h_0=h_c$ (obtained taking $\phi_0=-\pi/2$) 
and different frequencies: a power-law decay like $1/s$ is confirmed in agreement with Eq.~\eqref{g_critico:eqn}.
(Bottom) The critical case $h_{\rm i}=h_c=1$ for $h_0=0$ and $\phi_0=0$.
We see a $1/s^b$ decay with $b=3$ or $1$, depending on the Floquet spectrum being resonant ($2=l\omega_0$) or not in $k=0$.
}
\label{asymptotica-critical:fig}
\end{figure}
{\it Case c) $h_{\rm i}=h_c$:} 
When the initial field is critical, $E_k$ is gapless and the behaviour of $G_\infty(is)$ is power-law rather than exponential.
The upper panel of Fig.~\ref{asymptotica-critical:fig} illustrates several cases with $h_{\rm i}=h_c$, 
for a driving of the form $h(t) = h_0 + A\cos(\omega_0 t+\phi_0)$ with 
$h_0$ such that the Floquet spectrum is resonant with $l=0$, i.e., $h_0=h_{\rm i}=h_c$ (obtained for $\phi_0=\pm\pi/2$), 
showing a clear power-law decay of the form $1/s$:
\begin{equation} \label{g_critico:eqn}
  G_\infty(is)\simeq\nep^{Lg_\infty}\Big(1+\frac{D}{s}\Big) \;.
\end{equation}
(Remarkably, this $1/s$ decay is valid also for $\omega_0$ such that there is coherent destruction of tunnelling, 
for instance when $\omega_0=0.3623\ldots$, where $J_0(2A/\omega_0)=0$.)
Correspondingly, we predict a step-singularity in $P_{\infty}(W)$ 
\begin{equation}
  P_{\infty}(W)\simeq\nep^{Lg_\infty} \Big( \delta(W)+D\;\theta(W) \Big) \;,
\end{equation}
where $\theta$ is the Heaviside function.
%
We notice that in Ref.~\cite{Pietro_PRE} the authors find a very similar formula for $P_t(W)$ (when the time
$t$ is finite) in an Ising chain undergoing a generic (non-necessarily periodic) driving which ends at the critical point. 
In our case, we consider the asymptotic behaviour and, thanks to periodicity, whenever $h_{\rm i}=h_c$ the system 
not only ends but also starts at the critical point.
It turns out that the case $h_{\rm i}= h_c$ is quite rich, and other power-laws are possible if the Floquet spectrum
is resonant at $k=0$ with non-zero values of the integer $l$ in Eq.~\eqref{reso_cond:eqn}.
For instance, the bottom panel of Fig.~\ref{asymptotica-critical:fig} shows cases with $h_{\rm i}=h_c=1$, $A=1$ and $h_0=0$:
if the Floquet spectrum in $k=0$ is resonant ($2=l\omega_0$, Eq.~\eqref{reso_cond:eqn})
then $\frac{\ln{G}_{\infty}(is)}{L}-g_\infty$ decays like $1/s^3$, otherwise it decays like $1/s$. 
A $1/s^3$ behaviour, $G_\infty(is)\simeq\nep^{Lg_\infty}\Big(1+\frac{D}{s^3}\Big)$ results into a mild quadratic
increase at small $W$
for the work distribution: $P_{\infty}(W)\simeq\nep^{Lg_\infty} \Big( \delta(W)+\theta(W) D W^2/2 \Big)$.

\section{Finite temperature results: irreversible entropy generation} \label{Sec6}
One can also calculate the average work performed in the finite temperature case, where the initial state
is an equilibrium Gibbs state at a finite temperature $k_BT=\beta^{-1}$. 
Using Eq.~\eqref{logG_Ising:eqn} for $\ln G_{n\tau}(u)$, and the fact that the average work performed in a time $\tf=n\tau$ 
is given by $-i\partial \ln G_{n\tau}(u)/\partial u|_{u=0}$, one can readily arrive at the expression:
\begin{equation} \label{W_n_T:eqn}
\langle W \rangle_{n\tau} = 4L \int_{0}^{\pi} \! \frac{\ud k}{2\pi}\;  \left(1-\cos{(2\mu_k n\tau)}\right)
\left|r_k^+\right|^2\left|r_k^-\right|^2 E_{k} \tanh{(\beta E_k)} \;, 
\end{equation}
which generalizes Eq.~\eqref{Work_infty_T=0:eqn} to finite $T$ and finite $n$.
When $n\to \infty$, the rapidly oscillating term $\cos{(2\mu_k n\tau)}$ gives a contribution that averages to zero, and we get:
\begin{equation} \label{W_infty_T:eqn}
\langle W \rangle_{\infty} = 4L \int_{0}^{\pi} \! \frac{\ud k}{2\pi} \; \left|r_k^+\right|^2 \left|r_k^-\right|^2 E_{k} \tanh{(\beta E_k)} \;. 
\end{equation}
The statistical nature of the work for a finite system --- evidenced by Eq.~\eqref{eq_work_dis} --- calls for a second
law of thermodynamics written as $\langle W \rangle \geq \Delta F$, which relates the average work performed to the difference 
in equilibrium free energy $\Delta F$ corresponding to the initial and final parameter values: here the equality holds only for a quasi-static process. 
Calling $\langle W^{\rm irr} \rangle = \langle W \rangle - \Delta F$ the difference between $\langle W \rangle$ and $\Delta F$ 
--- and viewing it as the average irreversible work done ---, one can recast the second law in the form $\langle W^{\rm irr}\rangle \geq 0$. 
The heat transfer between the closed system and the bath being zero, the entire contribution to the entropy generation is in fact due to
$\langle W^{\rm irr} \rangle$, and one can define the irreversible entropy generated as 
$\Delta S^{\rm irr}=\beta\langle W \rangle^{\rm irr}$ \cite{jarzynski97,Deffner10}.

\begin{figure}[h]
\begin{center}
\includegraphics[height=9cm]{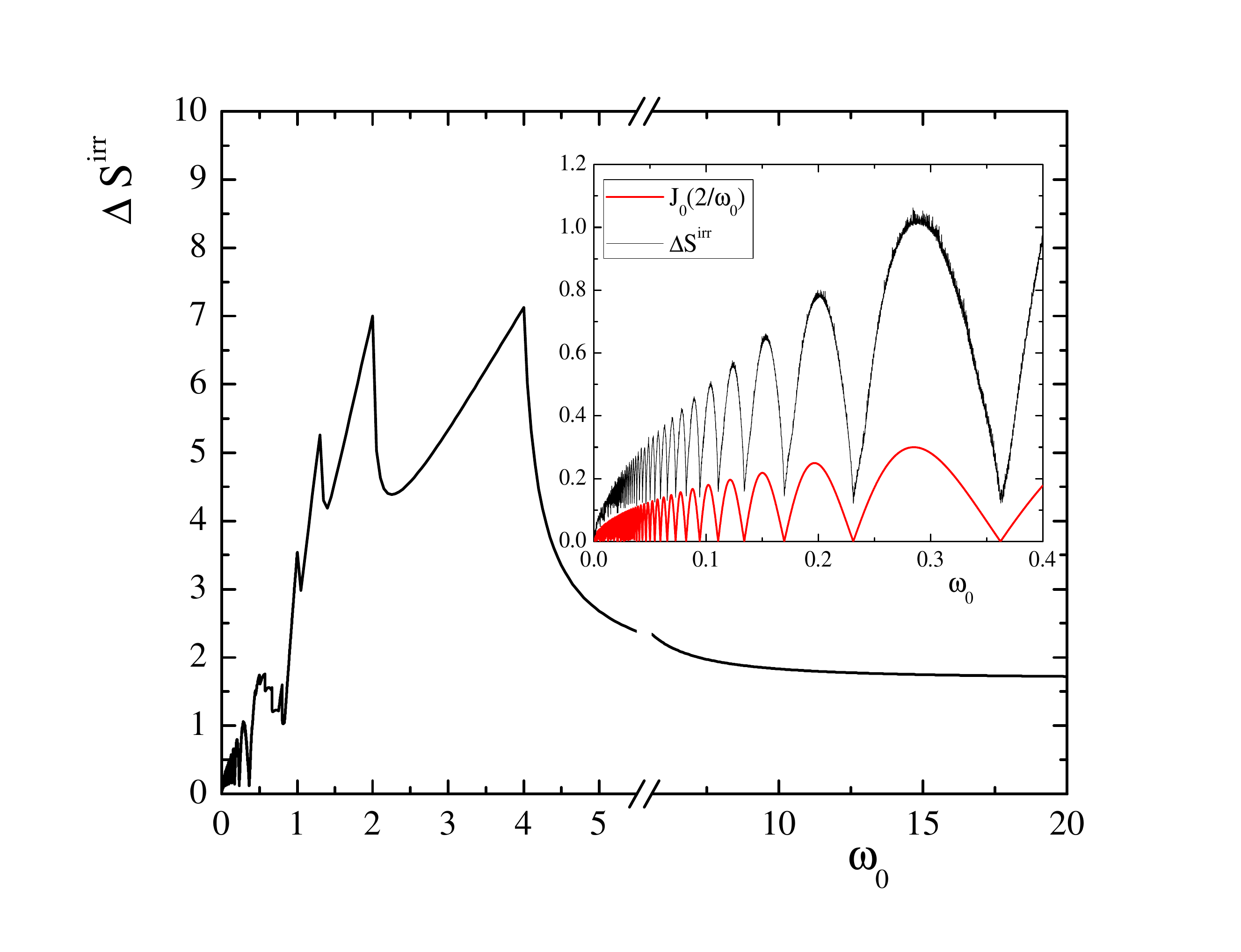} 
\end{center}
\caption{Figure showing that for a periodic driving protocol $h(t)=1 +\cos \omega_0 t$,  $\Delta S^{\rm irr}$  saturates to a steady state value in the asymptotic limit ($n \to \infty$) like the residual energy with similar dip and peak structure in the low  and intermediate $\omega_0$, respectively\cite{Russomanno_PRL12}.  In the main figure peaks occur at  $\omega_0 = 4/p$ while in the  inset we show that  dips in $\Delta S^{\rm irr}$ are located at $J_0(2/\omega_0)=0$ for small $\omega_0$. Here, $\beta=10$ and  $L=1000$. }
\label{Fig:finiteT}
\end{figure}

Since at stroboscopic times $t_n=n\tau$ the Hamiltonian returns to the original value $\Ham(0)$, so that $\Delta F=0$,
one can define an irreversible entropy increase, in the limit $n\to\infty$, as $\Delta S^{\rm irr}=\beta\langle W \rangle_{\infty}$.  
In Fig.~(\ref{Fig:finiteT}) we show that for a driving protocol $h(t) =1 + \cos (\omega_0 t)$,  
$\Delta S^{\rm irr}$ indeed saturates to a steady state value, like the residual energy \cite{Russomanno_PRL12}, displaying a sequence of well defined dips and peaks: In the small $\omega_0$ regime, there are dips at certain frequencies for which $ J_0(2A/\omega_0)=0$, a consequence of the coherent destruction of tunnelling \cite{Grossmann_PRL91}. 
In the intermediate range, on the contrary, one finds peaks at $\omega_0=4/p$, with $p$ integer, due to quasi-degeneracies
in the Floquet spectrum \cite{Russomanno_PRL12}.

\section{Conclusion}  \label{Sec7}
In conclusion, we have studied a periodically driven transverse-field Ising model and analyzed the behaviour 
of the  stroboscopic characteristic function $G_{n\tau}(is)$, and hence the stroboscopic work 
distribution function $P_{n \tau} (W)$, after $n$ complete periods of driving. 
Our study establishes that, in the thermodynamic limit,  $P_{n\tau}$ indeed converges for $n\to \infty$ towards
a well defined steady state value $P_{\infty}(W)$ which reproduces the exact asymptotic value of the first cumulant $\langle W \rangle_{\infty}$ 
(i.e., the asymptotic value of the average work performed on the system) derived earlier \cite{Russomanno_PRL12}. 
In the limit $s \to \infty$, on the other hand, $G_{n\tau}(is)$ reduces to the stroboscopic dynamical fidelity \cite{Sharma_EPL14}.  

For large $s$, we are able to provide asymptotic analytical expressions for $G_{\infty}(is)$ and, by means of inverse Laplace transforms,
we can derive corresponding expressions describing the small-$W$ behaviour of $P_{\infty}(W)$. 
The small-$W$ properties of $P_\infty(W)$ depend strongly on the fact that there is a static critical point $h_c$, and 
any periodic driving induces further non-equilibrium critical points where the gap in the Floquet spectrum closes up. 
This finding is in line with the study reported in Ref.~\cite{Batisdas_PRA12}, where, however, the relevant regime was one
of large-amplitude driving at large frequencies, and a rotating wave approximation was appropriate. 
Here, on the contrary, the exact resonances we find reside at low frequencies: for a fixed average field $h_0$, at $\omega_0=2|h_0-h_c|/l$. 
In any case, the form of the singularity in the work distribution turns out to be a useful detector of such non-equilibrium phase transitions.
According to the way the external periodic driving field relates to these critical points we can observe different phenomena. 
The time-averaged value of the field $h_0$ and its initial value $h_{\rm i}$ happen to be crucial.
Whenever $h_{\rm i}$ is different from the {\em static} critical point $h_c$ and $h_0$ differs from any
{\em non-equilibrium} critical point (i.e., $2|h_0- h_c|\neq l\omega_0$),
the asymptotic $P_{\infty}(W)$ is characterized by a universal edge singularity whose position $W_{\rm th}=2|h_{\rm i}-h_c|$
depends only on the value of $h_{\rm i}$. 
The information about the specific protocol appears only in the strength of the singularity. 
The low-energy behaviour of $P_{\infty}(W)$ is essentially determined by the lowest excitation modes and the corresponding
energy gap. If $h_{\rm i}\neq h_c$ but $h_0$ is dynamically critical (i.e., $2|h_0- h_c|= l\omega_0$ for some integer $l$), 
the form of this singularity changes, and becomes a Dirac delta derivative. 
A completely different behaviour emerges if the initial field is critical, $h_{\rm i}=h_c$: in this case, the cumulant generating
function (more precisely, $\frac{\ln{G}_{\infty}(is)}{L}-g_\infty$) has a power-law decay, as $1/s$, resulting
in a step-function contribution $P_{\infty}(W)$, but other decays, as $1/s^3$, can also be seen if the Floquet spectrum is resonant
with $l\neq 0$. 
Generalizing our investigation to the finite temperature case, we have shown that the irreversible entropy 
$\Delta S^{\rm irr}$, obtained using the first cumulant of the finite temperature characteristic function, also synchronizes with the periodic 
driving for $n \to \infty$ and converges to a steady state value for large $\omega_0$. 

Summarizing, we see a strong relationship between the features of $P_\infty(W)$, the existence
of a critical point and the way the driving field relates to it. This work is a first step towards the application of time-periodic
probes to understand the existence of a quantum phase transition by looking at the work distribution function. 
In this sense, we are generalizing the very interesting works in Refs.~\cite{Silva_PRL08, Gambassi_11:preprint,Gambassi12,Paraan_PRE09}, 
which refer to the case of a sudden quench. 
In this perspective, it is also interesting to see if it is possible to induce non-equilibrium phase
transitions in systems without static transitions and how this influences the work statistics. 
Another possible direction is to see how the quantum driven system being regular or ergodic 
(\cite{Russomanno:preprint,Lazarides_PRE14,Emanuele_2014:preprint,Abanin_AP15}) influences the work statistics. 
A lot of work still remains to be done, starting from the consideration of other 
tractable cases, like the Dicke model~\cite{Paraan_PRE09}. 

\appendix
%
%
\section{} \label{App:convergence}
In this appendix we show how the stroboscopic cumulant generating function tends towards the stationary value given by Eq.~\eqref{logG-asymp}.
Our first step is to expand the logarithm in Eq.~\eqref{logG_final_IsingT0:eqn}, which we report here for the reader's convenience
\begin{equation} \label{logG_final_IsingT0bis:eqn}
\frac{\ln G_{n\tau}^{T=0}(is)}{L} = \int_{0}^{\pi} \! \frac{\ud k}{2\pi} \; 
\ln \Big\{ 1- \frac{2q_k}{1+q_k} \sin^2{(\mu_k n\tau)} (1-\nep^{-2s E_{k}}) \Big\} \;.
\end{equation}
To that purpose, we first show that the second term inside the logarithm is $<1$. 
We know that $\sin^2(n\mu_k \tau)\leq 1$ and $|1-\nep^{-2sE_{k}}|< 1$ whenever $s>0$ and $s\neq \infty$. 
As for the overall prefactor, we notice that
\begin{equation} \label{xi_k:eqn}
  \xi_k\equiv \frac{2q_k}{1+q_k} = 4\left|r_k^+\right|^2 \left|r_k^-\right|^2 = 4 |r_k^+|^2 \left(1-|r_k^+|^2\right) \leq 1 \;,
\end{equation}
where the value $\xi_k=1$ is obtained for $|r_k^+|^2=1/2$.
%
%
Hence, the expansion of the logarithm is certainly possible for all $s<\infty$.
%
%
%
Defining 
\begin{equation} \label{defi_xi}
\xi_k(s) \equiv \frac{2q_k}{1+q_k} \left(1-\nep^{-2s E_{k}}\right) = \xi_k \left(1-\nep^{-2s E_{k}}\right) \;,
\end{equation}
we have:
\begin{equation} \label{logG-nice2}
  \frac{\ln G_{n\tau}^{T=0}(is)}{L} = -\sum_{m=1}^\infty \int_0^\pi\! \frac{\ud k}{2\pi}\; \frac{\xi_k^m(s)}{m} 
\sin^{2m} \left(\mu_k n\tau\right) \;,
\end{equation}
where we have exchanged the integral and the sum over $m$, due to the dominated convergence theorem. 
We then write a binomial expansion of the sine term in terms of exponentials:
\begin{equation}  \label{sin-exp}
  \sin^{2m} \left(\mu_k n\tau\right) = \frac{(-1)^m}{4^m} \sum_{j=0}^{2m} \binom{2m}{j} (-1)^j\nep^{2i(m-j)\mu_kn\tau} \;.
\end{equation}
%
The sum over $j$, for each $m$, has a finite number of terms: there is no problem in exchanging the integral over $k$ with this sum. 
We now observe that the  $j\neq m$ terms contain rapidly oscillating factors and vanish in the limit $n\to\infty$, 
thanks to the Riemann-Lebesgue lemma and the smoothness of the factors $\xi_k^m$. 
Hence, in the limit $n\to\infty$, we retain only the $j=m$ terms and write
\begin{equation} \label{logG-bellissimo}
  \lim_{n\to\infty}\frac{\ln G_{n\tau}^{T=0}(is)}{L} = -
  \int_0^\pi\! \frac{\ud k}{2\pi}\; \sum_{m=1}^\infty \frac{1}{4^m} \binom{2m}{m}\frac{\xi_k^m(s)}{m} \;.
\end{equation}
%
We can write this expression in a closed form, by defining 
\[ f(\xi)\equiv \sum_{m=1}^\infty\frac{1}{4^m}\binom{2m}{m}\frac{\xi^m}{m} \] 
and noticing that:
\begin{equation}
  \frac{\ud}{\ud\xi}f(\xi)=\frac{1}{\xi}\left(2\sqrt{\xi}\frac{\ud}{\ud\xi}\arcsin\left(\sqrt{\xi}\right)-1\right)\,,
\end{equation}
which can be integrated to give
\begin{eqnarray} \label{fixi:eq}
  f(\xi) &=& \int_0^\xi\frac{1}{\xi'}\left[\frac{1}{\sqrt{1-\xi'}}-1\right]\ud\xi'=2\int_0^{\arcsin\sqrt{\xi}}\tan\left(\frac{\eta}{2}\right)\ud\eta\nonumber\\
         &=&- 4 \Big[\ln\left(\cos\eta'\right)\Big]_0^{\frac{1}{2}\arcsin\sqrt{\xi}} = -2\ln\left[\frac{1}{2}\left(1+\sqrt{1-\xi}\right)\right]\,.
\end{eqnarray}
(In the last steps we have substituted $\xi=\sin^2(\eta)$ and $\eta'=\eta/2$.) 
In conclusion, we arrive at the desired result:
\begin{equation}   \label{logG_n_asympt}
  \lim_{n\to\infty} \frac{\ln G_{n\tau}^{T=0}(is)}{L} = 2 \int_0^\pi\! \frac{\ud k}{2\pi}\; \ln\left[\frac{1+\sqrt{1-\xi_k(s)}}{2}\right]
\end{equation}
where $\xi_k(s)$ is given in Eq.~\eqref{defi_xi}.

%
\section{} \label{App:low_energy}
We analyze here the asymptotic large-$s$ behaviour of $G_{\infty}^{T=0}(is)$. 
We start by assuming that the initial transverse field is not critical, $h_{\rm i}\neq h_c=1$, so that there is 
a gap in the spectrum of the initial Hamiltonian, $E_k\ge |h_{\rm i}-1|>0$.
We need this condition because we would like to expand Eq.~\eqref{logG_n_asympt} to lowest order in $\nep^{-sE_k}$ and this 
is possible, provided $E_k>0$ and $s\gg 1/|h_{\rm i}-1|$.
With the previously defined shorthand $\xi_k\equiv\xi_k(s\to\infty)=2q_k/(1+q_k)$ (which is a positive quantity in $[0,1]$) we can rewrite
Eq.~\eqref{logG_n_asympt} as
\begin{equation}   \label{logG_n_asympt_rewritten}
\frac{\ln G_{\infty}^{T=0}(is)}{L} = 2 \int_0^\pi\! \frac{\ud k}{2\pi} \; \ln\left[\frac{1+\sqrt{1-\xi_k+\xi_k\nep^{-2sE_k}}}{2}\right] \;.
\end{equation}
Expanding this to first order in $\nep^{-2sE_k}$ (with the assumption $\xi_k\neq 1$, that is $|r_k^+|^2\neq 1/2$, 
see comments below) we find:
\begin{equation}   \label{logG_n_asympt_e}
\frac{\ln G_{\infty}^{T=0}(is)}{L} \simeq \frac{\log G_{\infty}(is\to\infty)}{L} + 
              \int_0^\pi\! \frac{\ud k}{2\pi}\; \frac{\xi_k}{\sqrt{1-\xi_k}\left(1+\sqrt{1-\xi_k}\right)}\nep^{-2s E_k} \;.
\end{equation}
%
%
%
The expansion is questionable whenever there are $k$-points such that $\xi_k=1$.
We will see below that this is indeed the case at $k=0$ whenever the field oscillates around a non-equilibrium critical value given
by Eq.~\eqref{reso_cond:eqn} and the Floquet spectrum is resonant in $k=0$.
Even restricting ourselves to non-resonant cases, we would possibly find $k$-points where $\xi_{k_0}=1$, but
this time with $k_0>0$. 
This would still seem to be an issue at first glance: indeed, near these points we would expand
$\xi_k$ quadratically, $\xi_k=1-\lambda^2(k-k_0)^2+\mathcal{O}((k-k_0)^4)$, and we would therefore have, 
in the integrand of Eq.~\eqref{logG_n_asympt_e}, some logarithmic singularities originated by terms of the form $1/\left(k-k_0\right)$. 
In the neighborhood of these points where $\xi_{k_0}=1$, however, a different expansion is more appropriate. 
Considering one such $k_0$ where $\xi_{k_0}=1$, we can write the contribution to $\ln G_{\infty}^{T=0}(is)/L$ from the 
neighborhood of such a point, see Eq.~\eqref{logG_n_asympt_rewritten}, as
$$
\sim 2 \int_{k_0-\epsilon}^{k_0+\epsilon}\! \frac{\ud k}{2\pi}\; \ln\left[\frac{1+{\nep^{-sE_k}}}{2}\right]
\sim -\frac{2\ln 2}{\pi}\epsilon + {\nep^{-2sE_{k_0}}}\int_{-\epsilon}^{\epsilon}\frac{\ud k'}{2\pi}\nep^{-2sv_{k_0}k'}
$$
where we have approximated $\xi_k\simeq 1$, expanded the logarithm and introduced the group velocity 
$v_{k_0}=\left.\partial E_k/\partial k\right|_{k_0}$. 
This is indeed a convergent expression. More importantly, when $k_0\neq 0$ the singularities needing such a special treatment 
are in a region where the integrand is exponentially smaller, due to the prefactor $\nep^{-2sE_{k_0}}$,  
then the main contribution coming from the $k=0$ region, which we are now going to analyze. 


%
To proceed with the case $h_{\rm i}\neq h_c=1$, we expand $E_k$ up to second order in $k$
\begin{equation} \label{Ek:eqn}
E_k = |h_{\rm i}-1| + \frac{{h_{\rm i}}}{2|h_{\rm i}-1|} k^2 + \mathcal{O}(k^4) \;.
\end{equation}
Because we are assuming $|h_{\rm i}-1|>0$ and $s\gg1/|h_{\rm i}-1|$, only the $k$ such that $k\ll\sqrt{|h_{\rm i}-1|/h_{\rm i}}$ 
contribute significantly to the integral in Eq.~\eqref{logG_n_asympt_e}: we would be willing, therefore, to approximate the integral with 
a Gaussian one, extending the upper integration limit to $\infty$ and expanding the factor multiplying $\nep^{-2sE_k}$ to
the lowest order in $k$. But here the $k=0$ point plays a tricky role. 
For a generic periodic driving where $h(t)$ oscillates around the average value $h_0$ one quickly realizes,
by focusing on the evolution operator $\mathbb{U}_k(\tau,0)$ for modes with a small $k$ (as detailed in \ref{App:small_k}) that
there are two cases:
%
\begin{itemize}
\item[\em i)] The Floquet spectrum is resonant in $k=0$ whenever
\begin{equation}
  2|h_0-h_c| = l\omega_0 \quad\quad \textrm{for some positive integer}\; l 
\end{equation}
and then $\mu_k^{\pm}\propto \pm k$ and 
\begin{equation} \label{h01:eqn}
|r_k^+|^2=\frac{1}{2} - \frac{\beta k}{2}  + O(k^2)
\end{equation}
hence $\xi_0=1$ and the expansion in Eq.~\eqref{logG_n_asympt_e} is inappropriate.
As detailed in \ref{App:small_k}, for a sinusoidal driving of the form $h(t)=h_0+A\cos(\omega_0 t+\phi_0)$ there are special situations
where there is coherent destruction of tunnelling (CDT)~\cite{Hausinger_PRA10,Grossmann_PRL91}: if $J_l(2A/\omega_0)=0$, for the value of $l$ that
realizes the resonance, then a gap is opened in the Floquet spectrum and $\xi_0=0$, posing no problem with Eq.~\eqref{logG_n_asympt_e}.
\item[\em ii)] If, on the contrary, the Floquet spectrum is not resonant in $k=0$, then
\begin{equation} \label{h02:eqn}
|r_k^+|^2 = \frac{\alpha^2k^2}{4} + O(k^4) \;.
\end{equation}
hence $\xi_0=0$, again posing no problem with Eq.~\eqref{logG_n_asympt_e}.
%
\end{itemize}

So, restricting our consideration to non-critical initial fields, $h_{\rm i}\neq h_c$, and non-resonant Floquet spectrum,
and performing the appropriate Gaussian integral emerging from Eq.~\eqref{logG_n_asympt_e}, we finally arrive at:
\begin{equation} \label{app1:eqn}
\frac{\ln G_{\infty}^{T=0}(is)}{L} \simeq g_\infty + 
                      \frac{\alpha^2}{16\sqrt{\pi}} \left(\frac{|h_{\rm i}-1|}{h_{\rm i} s}\right)^{3/2} \nep^{-2s|h_{\rm i}-1|} + \cdots \;,
\end{equation}
where we assumed $s\gg1/|h_{\rm i}-1|$, while $\alpha$ is the constant appearing in the quadratic expansion of
$|r_k^+|^2$, see Eq.~\eqref{h02:eqn}.
Since $s$ is large, we can equivalently recast this equation as
\begin{equation}  \label{Gus}
  G_{\infty}^{T=0}(is) \simeq \nep^{Lg_{\infty}} \left( 1 +\frac{aL}{{s}^{3/2}} \nep^{-2s \left|h_{\rm i} - 1\right|} + \cdots \right) \;,
\end{equation}
where we have defined
\begin{equation} \label{a:eqn}
a \equiv \frac{\alpha^2}{16\sqrt{\pi}}\left(\frac{|h_{\rm i}-1|}{h_{\rm i}}\right)^{3/2} \;.
\end{equation}
Performing the inverse Laplace transform, it is not difficult to show that the $P(W)$ associated to Eq.~\eqref{Gus} is:
\begin{equation}\label{PW_asymp}
P_{\infty}(W) \simeq \nep^{Lg_{\infty}} \left( \delta(W) + \frac{2aL}{\sqrt{\pi}} \sqrt{W-2\left|h_{\rm i}-1\right|} 
   \; \theta\left(W-2\left|h_{\rm i}-1\right|\right)  + \cdots \right) \;.
\end{equation}
So, our theory predicts a square-root edge singularity in the asymptotic work distribution at a precise value 
$W_{\rm th}=2\left|h_{\rm i}-1\right|$ of $W$, i.e., simply the initial {\em gap} of the system, a value 
which is totally independent of the details of the periodic protocol (and even of the frequency), which only enter into the prefactor $a$. 

%
Finally, let us briefly consider the remaining cases where the previous Gaussian analysis fails. 
There are, essentially, two cases left:
\begin{enumerate}
\item \emph{Case $h_{\rm i}\neq h_c$ and resonant Floquet spectrum.}
The starting point of this case is Eq.~\eqref{logG_n_asympt_rewritten}, with $E_k$ given by Eq.~\eqref{Ek:eqn} but with  
the complication that $\xi_{k\to 0}=1$ in the vicinity of $k=0$. 
An analysis of the singularity emerging shows that the leading term for large $s$ is now given by:
\begin{equation}  \label{lnG_h01:eqn}
  \frac{\ln G_{\infty}^{T=0}(is)}{L} \simeq g_\infty \;+\; a_c \, s \, \nep^{-2 s |h_{\rm i}-1|} \; + \cdots \;.
\end{equation}
%
Exponentiating, we can equivalently rewrite:
\begin{equation}  \label{G_h01:eqn}
  G_{\infty}^{T=0}(is) \simeq \nep^{Lg_{\infty}}  \left( 1 \; + \; a_c \, L s \; \nep^{-2 s |h_{\rm i}-1|} + \cdots \right) \;. 
\end{equation}
The resulting $P(W)$, after inverse Laplace transform, has a very singular contribution proportional to a Dirac's delta derivative: 
\begin{equation} \label{Pinf-h0=1}
P_{\infty}(W) \simeq \nep^{Lg_{\infty}} \Big( \delta(W) \;+\; a_c \, L \; \delta'(W-2\left|h_{\rm i}-1\right|) 
                     \; \theta\left(W-2\left|h_{\rm i}-1\right|\right) \;+ \cdots \Big) \;
\end{equation}

\item \emph{Cases with $h_{\rm i}=h_c$}. 
Whenever the initial field $h_{\rm i}$ coincides with the critical field $h_c$ the initial spectrum $E_k$ is {\em gapless},
and this changes completely the large-$s$ behaviour of $G_{\infty}^{T=0}(is)$, from exponential to power-law.
The scenario is quite rich.  For instance, when $h_0=h_c$ the behaviour is of the form
\begin{equation} \label{g_critico-app:eqn}
  G_\infty(is)\simeq\nep^{Lg_\infty}\Big(1+\frac{D}{s}\Big) \;,
\end{equation}
leading to a small-$W$ probability distribution
\begin{equation}
  P_{\infty}(W)\simeq\nep^{Lg_\infty} \Big( \delta(W)+D\;\theta(W) \Big) \;,
\end{equation}
where $\theta$ is the Heaviside function. But this does not exhaust all the possibilities: when $h_0=0$ we find that
the leading asymptotics of $G_\infty(is)$ is $1/s^3$ rather than $1/s$ whenever the Floquet-resonant condition Eq.~\eqref{reso_cond:eqn}
is fulfilled. A thorough study of the gapless scenario is left to future studies.

\end{enumerate}

\section{} \label{App:small_k}
In this appendix we prove the statements leading to the resonance condition Eq.~\eqref{reso_cond:eqn}
and the related small-$k$ expansions Eqs.~\eqref{h01:eqn} and \eqref{h02:eqn}. 
Let us start with the resonance condition. The argument is very similar to the one reported in the Supplemental Material of 
Ref.~\cite{Russomanno_PRL12}. Let us consider Eq.~\eqref{formaggioska:eqn} with a generic time-periodic $h(t)$ 
with period $\tau=2\pi/\omega_0$ and apply to it the time-dependent rotation $\mathbb{V}_k(t)=\exp\left(-if(t)\sigma_z\right)$,
where $f(t)\equiv \int_0^t\ud t'\left(h(t')-h_0\right)$.
%
%
The matrix $\mathbb{H}_k(t)$ in the new representation has the form
\begin{equation}  \label{trans_form_H:eqn}
  \widetilde{\mathbb{H}}_k(t)=
       \mathbb{V}_k^\dagger(t)\mathbb{H}_{k}(t)\mathbb{V}_k(t)
       -i\mathbb{V}_k^\dagger(t)\dot{\mathbb{V}}_k(t)=
        \left[ \begin{array}{cc}
           	h_0-\cos k 	&-i\sin k\,\nep^{2if(t)}\\
		i\sin k\,\nep^{-2if(t)}	&-h_0+\cos k \end{array} \right] \;.
\end{equation}
The unitary matrix $\mathbb{V}_k(t)$ has the remarkable property that 
$\mathbb{V}_k(\tau)=\mathbb{V}_k(0)=\boldsymbol{1}$.
Because of that, we can immediately write the time-evolution operator over one period $\tau$ in the original frame as
\begin{equation} \label{U_storto:eqn}
  \mathbb{U}_k(\tau,0)=\overleftarrow{\top}\nep^{-i\int_0^\tau \widetilde{\mathbb{H}}_k(t)\ud t}\,.
\end{equation}
In essence, the Hamiltonian in the rotated frame directly determines the form of the Floquet modes and quasi-energies,
obtained by diagonalizing $\mathbb{U}_k(\tau,0)$.
With a vanishing driving, the Floquet spectrum is given by the eigenenergies of the Hamiltonian folded in the first Brillouin zone
~\cite{Russomanno_PRL12}, therefore we have a resonance whenever $2E_k=l\omega_0$ for some integer $l$. 
For a finite external field $h(t)$, most of the resonances turn into avoided crossings, but for the modes with $k=0$ and $k=\pi$. 
Indeed, we can see from Eq.~\eqref{trans_form_H:eqn} that the field couples to $\sin k$, which vanishes at $k=0,\pi$. 
If we diagonalize Eq.~\eqref{U_storto:eqn} at those values of $k$, we easily find the single-particle Floquet quasi-energies as
$\mu_0^\pm=\pm|h_0-1|$ and $\mu_\pi^\pm=\pm|h_0+1|$. 
The resonances at $k=0$ are particularly important since, for $s\to\infty$ only the smallest values of $E_k$ matter 
(see Eqs.~\eqref{logG-asymp} and~\eqref{xi_k:def}), and these occur near $k=0$.
Folding the quasi-energy into the first Brillouin zone, we can see that the Floquet spectrum is resonant at $k=0$ when:
\begin{equation} \label{resonando:eqn}
2|h_0-1|=l\omega_0\quad\quad \textrm{for some positive integer}\; l \;.
\end{equation}
%
%
%

We now move to establishing Eqs.~\eqref{h01:eqn} and \eqref{h02:eqn}. For this purpose we have to consider that 
the overlap factors $|r_k^\pm|^2=|\langle \phi_{k}^{\pm}(0) | \psi_{k}^{\rm gs}\rangle|^2$ are obtained from the 
Floquet modes $|\phi_{k}^{\pm}(0)\rangle$, which are the eigenstates of an Hermitian operator, the Floquet Hamiltonian $\mathbb{H}_{k}^{F}$,
defined as
\begin{equation} \label{defi_HF:eqn}
  \nep^{-i\tau \mathbb{H}_{k}^{F}} \equiv \overleftarrow{\top}\nep^{-i\int_0^\tau\ud t\,\mathbb{H}_k(t)}\,.
\end{equation}
(The Floquet quasi-energies are the eigenvalues of $\mathbb{H}_{k}^{F}$.)
From the above discussion, see Eqs.~\eqref{trans_form_H:eqn} and~\eqref{U_storto:eqn}, we immediately see that
$$
  \mathbb{H}_{k=0}^{F} = \left(\begin{array}{cc}
                              h_0-1&0\\
                                 0& 1-h_0
                              \end{array}\right) \;.
$$
Because the Floquet quasi-energies are defined up to translations of an integer number of $\omega_0$, this Hamiltonian is equivalent to
\begin{equation} \label{HF0:eqn}
  \mathbb{H}_{k=0}^{F} = \left(\begin{array}{cc}
                              \widetilde{h}&0\\
                                 0& -\widetilde{h}
                              \end{array}\right)\,,
\end{equation}
where $\widetilde{h}=(h_0-1)-l\omega_0/2$ is $h_0-1$ folded in the first Brillouin zone $[-\omega_0/2,\omega_0/2]$. 
We see that $\widetilde{h}=0$ whenever there is a resonance, i.e., Eq.~\eqref{resonando:eqn} is valid. 
Eq.~\eqref{defi_HF:eqn} shows that $\mathbb{H}_{k}^F$ must have the same symmetry properties of $\mathbb{H}_k(t)$. 
We see, from Eq.~(7), that the symmetry relation 
\begin{displaymath}
  \mathbb{H}_{-k}(t) = \sigma_z\mathbb{H}_{k}(t)\sigma_z
\end{displaymath}
is valid at all times. Because $\sigma_z$ is a time-independent unitary transformation, Eq.~\eqref{defi_HF:eqn}
implies that $\mathbb{H}_{-k}^F=\sigma_z\,\mathbb{H}_{k}^F\,\sigma_z$. 
Because of Eq.~\eqref{HF0:eqn}, and the relations $\Tr[\mathbb{H}_{k}(t)]=0$, %
$\sigma_z^3=\sigma_z$,\quad $\sigma_z\sigma_x\sigma_z=-\sigma_x$,\quad $\sigma_z\sigma_y\sigma_z=-\sigma_y$, we can 
write the following second-order-in-$k$ expansion~{
\footnote{The vanishing of the trace of $\mathbb{H}_{k}^F$ at any $k$ comes from the vanishing of the trace of $\mathbb{H}_{k}(t)$
and the formula $\Tr[\mathbb{H}_{k}^F]=\frac{1}{\tau}\int_0^\tau\Tr[\mathbb{H}_{k}(t)]\ud \tau$, which is a corollary of the
Liouville's theorem~\cite{Arnold:book}.} 
} of $\mathbb{H}_{k}^F$
\begin{myequation} \label{spanzone:eqn}
  \mathbb{H}_{{\rm small} \; k}^F = \left(\begin{array}{cc}
                              \widetilde{h}+a_z k^2 & (a_x- i a_y) k\\
                                 (a_x + i a_y) k & -\widetilde{h}-a_z k^2
                              \end{array}\right)+\mathcal{O}(k^3)\,.
\end{myequation}
In general the coefficients $a_x$, $a_y$ and $a_z$ are non-vanishing; they can vanish in some cases, giving rise to interesting phenomena 
which we will discuss later. Whenever the resonance condition Eq.~\eqref{resonando:eqn} is {\em not} fulfilled (hence $\widetilde{h}\neq 0$), the 
second-order-in-$k$ expansion of Floquet modes %
(expressed in the basis $\mathcal{B}=\{\ket{0},\,\opcdag{k}\opcdag{-k}\ket{0}\}$) is
\begin{myequation}
  \left[\ket{\phi_{{\rm small}\; k}^+}\right]_{\mathcal{B}} = 
        \left(\begin{array}{c}
                   1-\dfrac{1}{8}\dfrac{a_y^2+a_x^2}{\widetilde{h}^2} k^2\\
                      \\
                   \dfrac{1}{2}\dfrac{a_x+ia_y}{\widetilde{h}} k
              \end{array}\right)
\quad{\rm and}\quad 
  \left[\ket{\phi_{{\rm small}\; k}^-}\right]_{\mathcal{B}} = 
        \left(\begin{array}{c}
                   \dfrac{1}{2}\dfrac{a_x+ia_y}{\widetilde{h}} k\\
                     \\
                   1-\dfrac{1}{8}\dfrac{a_x^2+a_y^2}{\widetilde{h}^2} k^2
        \end{array}\right) \;,
%
\end{myequation}
which applies to the case $\widetilde{h}>0$; these two states should be exchanged if $\widetilde{h}<0$. 
Moving now to the initial Hamiltonian ground state $\ket{\psi_k^{\rm gs}}$, the diagonalization 
of Eq.~\eqref{hamil1} immediately gives (for $h_{\rm i}>1$) %
\begin{myequation} \label{ground:st}
  \left[\ket{\psi_{{\rm small}\; k}^{\rm gs}}\right]_{\mathcal{B}}=\left(\begin{array}{c} \dfrac{ik}{2(h_{\rm i}-1)}\\
                                                                                  \\
                                                                         1-\dfrac{k^2}{8(h_{\rm i}-1)^2}\end{array}\right) \;.
\end{myequation}
Hence, for the overlap $|r_k^+|^2$ we find
$$
  |r_k^+|^2 = |\langle \phi_{k}^{+} | \psi_{k}^{\rm gs}\rangle|^2 =\frac{1}{4}\alpha^2 k^2+\mathcal{O}(k^3)
       \quad{\rm with}\quad \alpha^2 \equiv\left(\frac{1}{|h_{\rm i}-1|}-\frac{a_y}{|\tilde h|}\right)^2+\left(\frac{a_x}{\widetilde{h}}\right)^2
$$
which is indeed Eq.~\eqref{h02:eqn}; this formula is valid also in the case $\widetilde{h}<0$ and $h_{\rm i}<1$. 
If $\widetilde{h}<0$ and $h_{\rm i}>1$, or $\widetilde{h}>0$ and $h_{\rm i}<1$, we find 
$|r_k^-|^2= \frac{\alpha^2}{4}k^2+\mathcal{O}(k^3)$, but the crucial ingredient determining Eq.~\eqref{logG_n_asympt} %
is identical, since $\xi_k\simeq \alpha^2 k^2$ in both cases (see Eq.~\eqref{xi_k:eqn}). 

For the resonant case $\widetilde{h}=0$ we find
\begin{displaymath}
  \left[\ket{\phi_{{\rm small}\; k}^+}\right]_{\mathcal{B}} = 
        \frac{1}{\sqrt{2}}\left(\begin{array}{c}
                   1+\dfrac{1}{2}\dfrac{a_z}{\sqrt{a_x^2+a_y^2}}k\\
                   \\
                   \dfrac{-a_x+ia_y}{\sqrt{a_x^2+a_y^2}}\left(1-\dfrac{1}{2}\dfrac{a_z}{\sqrt{a_x^2+a_y^2}} k\right)
              \end{array}\right)
\end{displaymath}
%
which (if $h_{\rm i}>1$) gives rise to 
\begin{displaymath}
  |r_k^+|^2 = \frac{1}{2}\left(1-\frac{a_z}{\sqrt{a_x^2+a_y^2}}k\right)+\mathcal{O}(k^2)\,.
\end{displaymath}
This is indeed Eq.~\eqref{h01:eqn} with $\beta = a_z/{\sqrt{a_x^2+a_y^2}}$. 
We notice that this formula is valid if $a_x-ia_y\neq 0$. 
This is generally true, up to special cases where there is coherent destruction of tunnelling (CDT) ~\cite{Hausinger_PRA10}. 
where $a_x=a_y=0$, and we fall back to Eq.~\eqref{h02:eqn}. 
In the Supplemental material of Ref.~\cite{Russomanno_PRL12} we show that, %
in the case of a sinusoidal driving $h(t)=h_0+A\cos(\omega_0t + \phi_0)$, CDT occurs if $h_0=1$ and $J_0(2A/\omega_0)=0$.
More in general, if the resonance condition Eq.~\eqref{resonando:eqn} is valid for $l\neq 0$, we can show, exactly with the same %
arguments used in Ref.~\cite{Russomanno_PRL12}, that there is CDT whenever $J_l(2A/\omega_0)=0$.
\ack 
We acknowledge discussions with E.~G. Dalla Torre. 
Research was supported by the Coleman-Soref foundation, by MIUR, through PRIN-2010LLKJBX\_001, 
by SNSF, through SINERGIA Project CRSII2 136287/1, and by the EU FP7 under grant agreement n. 280555, and the 
ERC Advanced Research Grant N. 320796  MODPHYSFRICT.  SS acknowledges CSIR, India
and AD acknowledges SERB, DST, India for financial support.
\bigskip
\bigskip 
%

\end{document}